\documentclass[pre,aps,floats,twocolumn,superscriptaddress,floatfix]{revtex4}
\usepackage{graphicx,amssymb,amsmath,ifthen,rotating}
\usepackage[active]{srcltx}
\begin{document}
\title{Molecular chains under tension:
Thermal and mechanical activation of statistically interacting extension and contraction particles}  
\author{Aaron C. Meyer}
\affiliation{
  Department of Physics,
  University of Rhode Island,
  Kingston RI 02881, USA}
\author{Yahya \"Oz}
\affiliation{
Fachgruppe Physik,
  Bergische Universit{\"{a}}t Wuppertal,
  D-42097 Wuppertal, Germany}
 \author{Norman Gundlach}
\affiliation{
Fachgruppe Physik,
  Bergische Universit{\"{a}}t Wuppertal,
  D-42097 Wuppertal, Germany}
  \author{Michael Karbach}
\affiliation{
Fachgruppe Physik,
  Bergische Universit{\"{a}}t Wuppertal,
  D-42097 Wuppertal, Germany}
        \author{Ping Lu}
\affiliation{
Department of Applied Science and Mathematics,
  Bluefield State College,
  Bluefield, WV 24701, USA} 
\author{Gerhard M{\"{u}}ller}
\affiliation{
  Department of Physics,
  University of Rhode Island,
  Kingston RI 02881, USA}

\begin{abstract}
This work introduces a methodology for the statistical mechanical analysis of polymeric chains under tension controlled by optical or magnetic tweezers at thermal equilibrium with an embedding fluid medium.
The response of single bonds between monomers or of entire groups of monomers to tension is governed by the activation of statistically interacting particles representing quanta of extension or contraction.
This method of analysis is capable of describing thermal unbending of the freely jointed or wormlike chain kind, linear or nonlinear contour elasticity, and structural transformations including effects of cooperativity.
The versatility of this approach is demonstrated in an application to double-stranded DNA undergoing torsionally unconstrained stretching across three regimes of mechanical response including an overstretching transition.
The three-regime force-extension characteristic, derived from a single free-energy expression, accurately matches empirical evidence.
\end{abstract}
\maketitle

%
\section{Introduction}\label{sec:intro}
%
Experimental, computational, and theoretical studies of mechanical and thermal responses of molecular chains to tension, torque, and environmental change continue to be confronted with new vistas and new challenges due to astonishing advances in nanoscale technology, notably single-molecule manipulation techniques \cite{APRW16, MBM+17, HGH02, BZS+17, BSG+03, BBS03}.
The work reported here and in two related studies \cite{mct2,mct3} aims to shed light on the statistical mechanics of molecular chains under tension and torque and in interaction with molecules of the embedding fluid medium by a methodology not hitherto applied to this field, yet well developed in other areas of one-dimensional structures.

The fundamental degrees of freedom for the statistical mechanical analysis in the present context are the bonds between the molecules that form chain structures.
We shall use, for the purpose of statistical mechanical modeling, the generic term \emph{link} for diverse yet specific constructs incorporating energetic and geometric attributes of either individual bonds between monomers or segments of monomers and the bonds between them in a polymeric chain.

\begin{figure}[b]
  \begin{center}
\includegraphics[width=50mm]{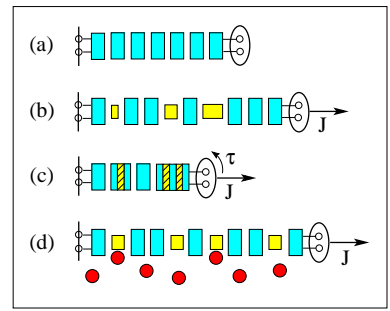}
\end{center}
\caption{(Color online) Schematic representations for a chain of $N=7$ monomers of (a) the reference state, (b) a state under tension with extension particles activated, (c) a state under tension and torque with twist-contraction particles activated, and (d) a state under tension with extension particles and contact particles activated.}
  \label{fig:figure1}
\end{figure}

The modifications of links in response to tension, torque, and contact with the embedding fluid are represented by the (thermal or mechanical) activation of statistically interacting quasiparticles.
Three basic types of particles are schematically illustrated in Fig.~\ref{fig:figure1}.
They involve particles carrying quanta of extension or contraction distances, twist angles, and contact energies.
Further types will be introduced as needed in particular applications.
We have already tested and used this methodology in a wide range of applications including collective modes with fractional statistics in magnetic chains \cite{LMK09}; effects of condensation in $\mathcal{D}$-dimensional quantum gases \cite{PMK07}; lattice gases with short-range and long-range couplings \cite{sivp}; granular matter jammed by contact, gravity, or centrifuge \cite{GKLM13, janac2}; and coil-helix transitions in polypeptides adsorbed to a water-lipid interface \cite{cohetra}.

All forms of bond modifications are discretized at a microscopic level as is common in other methods of statistical mechanical modeling.
The activation of extension particles produces quanta of incremental length under tension.
Likewise, quanta of incremental twist angle under torque are generated by the activation of twist particles.
The latter may be simultaneously associated with a length contraction, as indicated in Fig.~\ref{fig:figure1}(c).
Each particle species is assigned an activation energy which, in general, depends on tension and torque.
The versatility of this methodology makes it suitable for the interpretation of experimental data on elastic responses and conformational changes.

In-depth studies of the mechanics of macromolecules have been made possible by recent advances in single-molecule manipulation techniques, specifically the use of optical and magnetic tweezers on double-stranded (ds) DNA.
With such experiments, tension in the fN and pN ranges can be controllably applied. 
DNA has been observed to respond in a very complex way to forces and torques within the experimentally accessible windows \cite{APRW16, MBM+17, HGH02, BZS+17, BSG+03, BBS03, Wang97, Strick96, MN13}.
For a typical case of torsionally unconstrained stretching, at low tensions ($<$10 pN) entropically dominated thermal unbending is the principal response \cite{Bust94}. 
At stronger tensions (10-60 pN), it gives way to an elasticity that is increasingly enthalpic \cite{Cluzel96,Smith96,Wang97}. 
Then, at a critical tension (65-70 pN) a rather abrupt conformational change can be observed, where DNA rapidly extends by more than two thirds of its contour length \cite{Cluzel96, Smith96, BEP+12, BEP+14, KGB+13, BML+14}.

The generic mathematical structure of our approach is based on an idea of Haldane \cite{Hald91a} and has been developed by many contributors \cite{Wu94, Isak94, Anghel, NA14, LVP+08, copic, picnnn, pichs}.
The inner workings of the methodology are outlined in Sec.~\ref{sec:sip}.
Relevant aspects of the emerging taxonomy of particle species come to light in the process.
This paper then covers three areas of applications.
The first two pertain to enthalpic elasticity (Sec.~\ref{sec:for-ext}) and entropic elasticity (Sec.~\ref{sec:appe}) in a generic sense.
The last is an application to double-stranded DNA under torsionally unconstrained tension subject  both kinds of elasticity in sequence and then undergoing a structural transition (Sec.~\ref{sec:DNA-1}).

Further applications of the same methodology  are work in progress.
One study \cite{mct2} has its focus on the mechanical response of molecular chains to tension and torque including the formation of plectonemes.
A second extension \cite{mct3} investigates environmental effects to mechanical responses caused by the fluid medium in which the molecular chain is embedded. 
The effects include elastic softening, intercalation, hysteresis, and other forms of irreversibility.

%
\section{Statistically interacting particles}\label{sec:sip}
%
Molecular chains described as a sequence of links modified by tension, torque, thermal agitations, and contact with specific molecules of the embedding fluid lend themselves naturally to an analysis as a system of statistically interacting quasi-particles whose activation energies depend on tension, torque, and chemical potentials.
The statistical interaction is governed by combinatorial rules that are captured in Haldane's generalized exclusion principle, originally proposed in a quantum many-body context \cite{Hald91a}. 
This concept of statistical interaction as a tool in statistical mechanical modeling has proven useful for a broad field of applications \cite{LMK09, PMK07, sivp, GKLM13, janac2, cohetra, Wu94, Isak94, Anghel, NA14, LVP+08, copic, picnnn, pichs}.

\subsection{Generalized Pauli principle}\label{sec:pauli}
Consider a system of particles from species $m=1,\ldots,M$.
Placing one particle of species $m'$ into the system, $\Delta N_{m'}=1$, affects the number $d_m$ of open slots for the placement of particles from any species $m$.
This effect is encoded in the generalized Pauli principle \cite{Hald91a},
\begin{equation}\label{eq:1} 
\Delta d_m=-\sum_{m'=1}^Mg_{mm'}\Delta N_{m'}\quad :~ m=1,\ldots,M,
\end{equation}
and specified by a set of rational numbers $g_{mm'}$, named statistical interaction coefficients.

The reference state (pseudo-vacuum) contains no particles and is unique in the present context.
It has a definite capacity for placing particles from each species separately.
That accommodation capacity is specified by a set of (non-negative, rational) capacity constants, $A_m$, $m=1,\ldots,M$.
The number of distinct microstates with particle content $\{N_m\}$ is determined by the following binomial multiplicity expression \cite{Hald91a, LMK09, Wu94, Isak94, Anghel}:
\begin{subequations}\label{eq:2} 
\begin{align}\label{eq:2a} 
W(\{N_m\}) &=\prod_{m=1}^M\left(\begin{array}{c}
d_m+N_m-1 \\ N_m\end{array}\right), \\ \label{eq:2b} 
 d_m &=A_m-\sum_{m'=1}^M g_{mm'}(N_{m'}-\delta_{mm'}).
\end{align}
\end{subequations}

The energy of a microstate only depends on its particle content,
\begin{equation}\label{eq:3} 
E(\{N_m\})=E_0+\sum_{m=1}^M N_m\epsilon_m,
\end{equation}
where $E_0$ is the energy of the reference state and the $\epsilon_m$ are the particle activation energies.
Interaction energies between elementary degrees of freedom are built into attributes of the more complex particles in use here as will become more transparent in the applications below.

\subsection{Partition function}\label{sec:part}
Techniques of evaluating the partition function,
\begin{equation}\label{eq:4} 
Z=\sum_{\{N_m\}}W(\{N_m\})e^{-\beta E(\{N_m\})},\quad 
\beta\doteq\frac{1}{k_\mathrm{B}T},
\end{equation}
with ingredients (\ref{eq:2}) and (\ref{eq:3}) were developed and carried out on rigorous grounds for macroscopic systems at different levels of generality by Wu \cite{Wu94}, Isakov \cite{Isak94}, Anghel \cite{Anghel}, and others \cite{LMK09, PMK07,sivp}.
A convenient way to express the partition function is
\begin{equation}\label{eq:5} 
Z=\prod_{m=1}^M\big(1+w_m^{-1}\big)^{A_m},
\end{equation}
where the (real, positive) $w_m$ are solutions of the coupled nonlinear algebraic equations,
\begin{equation}\label{eq:6} 
e^{\beta\epsilon_m}=(1+w_m)\prod_{m'=1}^M \big(1+w_{m'}^{-1}\big)^{-g_{m'm}}.
\end{equation}

The average numbers $\langle N_m\rangle$ of particles from all species are the solutions, for given $\{w_m\}$, of the coupled linear equations,
\begin{equation}\label{eq:7} 
w_m\langle N_m\rangle+\sum_{m'=1}^Mg_{mm'}\langle N_{m'}\rangle =A_m.
\end{equation}
The configurational entropy,
\begin{subequations}\label{eq:8}
\begin{align}\label{eq:8a}
S(\{\langle N_m\rangle\}) &= k_\mathrm{B}\sum_{m=1}^M
\Big[\big(\langle N_m\rangle+{Y}_m\big)\ln\big(\langle N_m\rangle+{Y}_m\big) \nonumber \\
&-\langle N_m\rangle\ln \langle N_m\rangle -{Y}_m\ln {Y}_m\Big], \\ \label{eq:8b}
 {Y}_m &\doteq {A}_m-\sum_{m'=1}^Mg_{mm'} \langle N_{m'}\rangle,
\end{align}
\end{subequations}
inferred directly from (\ref{eq:2}), via $S=k_\mathrm{B}\ln W$ in combination with (\ref{eq:7}), is equivalent to the expression obtained via derivative from the free energy related to (\ref{eq:5}).

\subsection{Nesting of particles}\label{sec:nest}
A significant broadening in scope of this methodology resulted from its extension to particle nesting \cite{LVP+08, copic, picnnn,pichs,cohetra,sivp}.
The generalized Pauli principle (\ref{eq:1}), in its original version, was meant to be an exclusion principle, implying that $\Delta d_m\leq0$ if $\Delta N_{m'}>0$.
All interaction coefficients must then be non-negative, $g_{mm'}\geq0$.
The capacity constants $A_m$ in the integrated version (\ref{eq:2b}) of (\ref{eq:1}) must be positive and grow linearly with system size, which ensures that the free energy inferred from (\ref{eq:5}) is thermodynamically extensive.

Nesting includes particles that cannot be activated from the reference state but must be hosted by particles from other species.
The former exist on top or inside the latter.
Hosted particles may, in turn, be hosts to other particles.
The hierarchical hosting structure may have any number of levels.
The taxonomy proposed in Ref.~\cite{copic}, (provisionally) introduced the categories of \emph{compacts}, \emph{hosts}, \emph{hybrids}, \emph{tags}, and \emph{caps} for particle species to distinguish their roles in the hierarchy.

Particles that exist side by side in the pseudovacuum are named compacts if they have no hosting capacity and hosts if they do. 
Particles that must be hosted are named hybrids if they also have a hosting capacity and tags or caps if they do not.
Tags leave space for further tags on the same host or hybrid, caps do not.

Compacts and hosts have capacity constants $A_m>0$, whereas all hosted particle species have $A_m=0$.
Capacity for the placement of hybrids, tags, and caps must be created by the prior placement of hosting particles (hosts or hybrids).
This requires that the interaction coefficient $g_{mm'}$ between a species $m'$ that hosts species $m$ must be negative.
Here the \emph{exclusion} principle turns into an \emph{accommodation} principle: $\Delta N_{m'}>0$ yields $\Delta d_m>0$.
The partition function (\ref{eq:5}) is a product where each factor represents a species of compacts or hosts.
The hosted species, which have $A_m=0$, only contribute indirectly via the solutions of Eqs.~(\ref{eq:6}).

It will become evident that the nesting of particles is an important feature for the modeling of cooperative processes in which one conformation nucleates and grows out of another conformation.
Particle nesting plays a key part in several applications to molecular chains under tension and torque worked out here and in \cite{mct2, mct3}. Further evidence was previously presented in the context of the coil-helix transition in polypeptides \cite{cohetra}.

\subsection{Particles at two levels}\label{sec:2-lev}
\begin{figure}[htb]
  \begin{center}
\includegraphics[width=85mm]{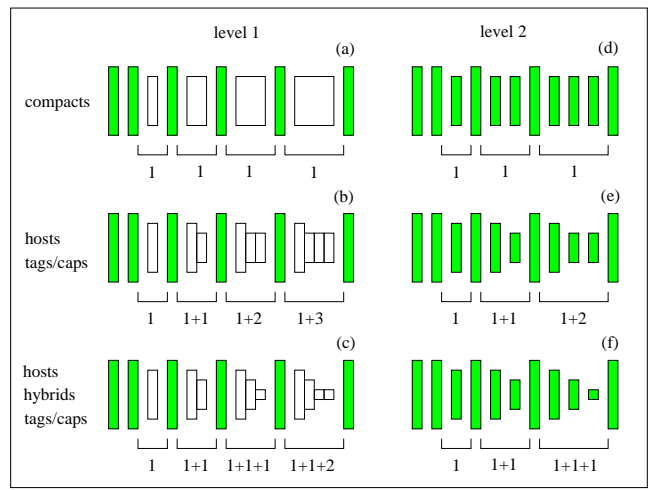}
\end{center}
\caption{Compact and nested particles populating single bonds between monomers at level 1 (left) or multiple monomers monomers at level 2 (right). 
Monomers are represented by shaded rectangles (six at level 1, eleven at level 2). 
Compacts are represented by rectangles of reduced height (top row).
Level-1 compacts are distinguished by open rectangles of different width.
Level-2 compacts are distinguished by groups containing different numbers of rectangles.
Nested particles (hosts, hybrids, tags) are represented by particles of one width but different heights. 
At level 1 they occupy (modified) bonds between adjacent monomers.
At level 2 they are made of groups of monomers with modified bonds.}
  \label{fig:figure6}
\end{figure}

A general taxonomy of statistically interacting particles -- still in the making -- will have to account for the distinction of two levels at which particles are placed in the system.
In the present context, particles at level 1 occupy single bonds between monomers  whereas particles at level 2 occupy one or several monomers including the bonds these monomers participate in.
Statistically interacting particles at both levels can be compact or nested.
In Fig.~\ref{fig:figure6} we show symbolic representations of these distinctive traits.

Microstate (a) contains four level-1 compacts and microstate (d) three level-2 compacts, all from different species.
Microstate (b) contains four level-1 hosts and microstate (e) three level-2 hosts. Hosts with one tag or one cap are marked \textsf{1+1}.
Hosts with two or three tags are marked \textsf{1+2}, \textsf{1+3}, respectively. 
In microstates (c) and (f) \textsf{1+1} means one host plus one hybrid, \textsf{1+1+1} means adding one tag or one cap, and \textsf{1+1+2} means adding a second tag to the first. 

The statistical mechanics of compact and nested particles at level 1 was worked out in great generality for a lattice-gas application \cite{sivp}.
The extension particles that we introduce in Secs.~\ref{sec:com-ext-par-1} and \ref{sec:nes-ext-par-1} belong to these types.
Compact and nested particles at level 2 were introduced to describe ferromagnetic domains and antiferromagnetic domain walls \cite{LVP+08,copic,picnnn,pichs}.
They were also employed to describe the coil-helix transition of a polypeptide \cite{cohetra}.

\subsection{Compact extension particles}\label{sec:com-ext-par-1}
A chain of $N$ monomers bound into a polymer of ${N-1}$ bonds responds to tension by extension while subjected to thermal agitation.
Bond modifications under tension are here described as the activation of compact level-1 extension particles [Fig.~\ref{fig:figure6}(a)].
We consider an arbitrary number $M$ of species and assign to each species an activation energy of the general form,
\begin{equation}\label{eq:9} 
\epsilon_m=\gamma_m-JL_m\quad :~ m=1,\ldots,M,
\end{equation}
where $J\geq0$ is the applied tension, $\gamma_m$ is an elastic-energy constant, and the $L_m$ are increments of length.

Each bond can either be in its original state or occupied by one compact (from any species).
The activation of a particle from species $m$ extends the end-to-end distance by $\Delta L=L_m>0$.
The chain also stretches out when a particle of size $L_m$ is replaced by a particle of size $L_{m'}>L_m$.
Conversely, the chain contracts if longer particles are replaced by shorter ones or if some bonds are deactivated altogether.
The dependence of $\epsilon_m$ on $J$ as specified has the consequence that with growing tension longer extension particles crowd out shorter ones.

The exact statistical mechanical analysis of level-1 compacts was previously worked out \cite{sivp}.
The combinatorics is governed by the multiplicity expression (\ref{eq:2}) with ingredients 
\begin{equation}\label{eq:10} 
A_m=N-1,\qquad g_{mm'}=\left\{
\begin{array}{ll} 1 &: m'\geq m, \\ 0 &: m'<m.
\end{array} \right.
\end{equation}
The energetic specifications (\ref{eq:9}) and combinatoric specifications (\ref{eq:10}) then set the table for the statistical mechanical analysis (\ref{eq:5})-(\ref{eq:7}).
The physically relevant exact solution of Eqs.~(\ref{eq:6}) reads
\begin{equation}\label{eq:11}
w_m=e^{\beta\epsilon_m}\left[1+\sum_{m'=1}^{m-1}e^{-\beta\epsilon_{m'}}\right] ~~:~ m=1,2,\ldots,M.
\end{equation}

All thermodynamic quantities of interest can be expressed via the functions,
\begin{equation}\label{eq:12} 
 B_{lk}(T,J)\doteq\sum_{m=0}^M (L_m/L_\mathrm{c})^l(\beta\epsilon_m)^ke^{-\beta\epsilon_m},
\end{equation}
with $\epsilon_0=0$ and $L_0=0$ implied and where $L_\mathrm{c}$ is some unit of excess length.
The partition function (\ref{eq:5}) inferred from the solution (\ref{eq:11}) has a very simple structure:
\begin{equation}\label{eq:45}
Z=\big[B_{00}\big]^{N-1}=\left[1+\sum_{m=1}^Me^{-\beta\epsilon_m}\right]^{N-1}.
\end{equation}
Note that the sum in (\ref{eq:45}) is over particle species, not over microstates as in (\ref{eq:4}).
The Gibbs free energy (per bond) inferred from (\ref{eq:45}) reads
\begin{equation}\label{eq:13} 
\bar{G}(T,J)=-k_\mathrm{B}T\ln B_{00}.
\end{equation}
Expressions for excess length, entropy, enthalpy, and internal energy (per bond) then follow directly:
\begin{equation}\label{eq:14} 
\bar{L}\doteq-\left(\frac{\partial\bar{G}}{\partial J}\right)_T =L_\mathrm{c}\frac{B_{10}}{B_{00}},
 \end{equation}
 \begin{equation}\label{eq:15}
 \bar{S}\doteq-\left(\frac{\partial\bar{G}}{\partial T}\right)_J 
= k_\mathrm{B}\left[\ln B_{00}+\frac{B_{01}}{B_{00}}\right],
\end{equation}
\begin{equation}\label{eq:16} 
\bar{H}=\bar{G}+T\bar{S},\quad \bar{U}=\bar{H}+J\bar{L}.
\end{equation}
$\bar{H}$ is the average energy of activated particles and $\bar{U}$ the average elastic energy in the system. 

Three response functions of thermal, mechanical, and mixed variety, named heat capacity, tensile compliance, and thermal expansivity, respectively, are expressed as follows:
\begin{equation}\label{eq:17} 
 \bar{C}_J\doteq T\left(\frac{\partial\bar{S}}{\partial T}\right)_J 
 =k_\mathrm{B}\left[ \frac{B_{02}}{B_{00}}-\left( \frac{B_{01}}{B_{00}}\right)^2\right],
\end{equation}
\begin{equation}\label{eq:18} 
\kappa_T\doteq \frac{1}{L_\mathrm{c}}\left(\frac{\partial\bar{L}}{\partial J}\right)_T 
=\frac{L_\mathrm{c}}{k_\mathrm{B}T}\left[\frac{B_{20}}{B_{00}}-\left(\frac{B_{10}}{B_{00}}\right)^2\right],
 \end{equation}
\begin{equation}\label{eq:19} 
 \alpha_J\doteq \frac{1}{L_\mathrm{c}}\left(\frac{\partial\bar{L}}{\partial T}\right)_J
=\frac{1}{T}\left[\frac{B_{11}}{B_{00}}-\frac{B_{10}B_{01}}{B_{00}^2}\right].
\end{equation}

The population densities of extension particles from each species as derived from (\ref{eq:7}) become
\begin{equation}\label{eq:20} 
\bar{N}_m\doteq \frac{\langle N_m\rangle}{N}=\frac{e^{-\beta\epsilon_m}}{B_{00}}\quad :~ m=0,1,\ldots,M,
\end{equation}
where $\bar{N}_0$ is the fraction of deactivated bonds.
An equivalent expression for the excess length (\ref{eq:14}) is
\begin{equation}\label{eq:77} 
\bar{L}=\sum_{m=0}^M\bar{N}_mL_m,
\end{equation}
and an equivalent expression for the entropy (\ref{eq:15}) can be inferred from (\ref{eq:8}) with appropriate scaling.

In most applications it is useful to assign excess lengths in the form,
\begin{equation}\label{eq:21} 
L_m\doteq mL_\mathrm{c}\quad :~ m=1,2,\ldots,M,
\end{equation}
to extension particles from species $m$. 
Each application is specified by a particular set of elastic-energy constants $\gamma_m$.
If we wish to restrict the permissible quanta of extension to a subset  $\{m_1,m_2,\ldots\}$, then we simply freeze out the remaining species by setting ${\gamma_m\to\infty}$ for $m\notin\{m_1,m_2,\ldots\}$.

\subsection{Nested extension particles}\label{sec:nes-ext-par-1}
Instead of \emph{compacts} we can employ \emph{hosts}, \emph{hybrids}, \emph{tags}, and \emph{caps} to cause the same bond modifications in response to tension.
These particles coexist on one and the same bond in nested combinations.
Here we limit the discussion to scenarios with one host species $(m=1)$, multiple hybrid species $(m=2,\ldots,M-1)$, and one species $(m=M)$ of tags or caps [Fig.~\ref{fig:figure6}(c)].
One significant extension of this scheme is introduced in Appendix~\ref{sec:appg}.

The case $M=2$ includes no hybrids.
Host and cap are equivalent to two species of compacts.
The first compact becomes the host without cap and the second compact becomes the host with cap.
Host and tags are equivalent to infinitely many compacts: host, host plus tag, host plus two tags etc.

The nested particles have activation energies of the general form (\ref{eq:9}).
Their combinatorics is governed by the multiplicity expression (\ref{eq:2}) with specifications
\begin{equation}\label{eq:a1} 
A_m = \left\{\begin{array}{ll}  \rule[-2mm]{0mm}{5mm}
N-1 &:~ m=1, \\  \rule[-2mm]{0mm}{5mm}
0 &:~ m=2,\ldots,M, 
\end{array} \right.
\end{equation}
\begin{equation}\label{eq:a2} 
g_{mm'} = \left\{\begin{array}{ll}  \rule[-2mm]{0mm}{5mm}
\delta_{m'm} &:~ m=1, \\  \rule[-2mm]{0mm}{5mm}
\delta_{m'm}-\delta_{m',m-1} &:~ m=2,\ldots,M-1, \\  \rule[-2mm]{0mm}{5mm}
-\delta_{m',m-1} &:~ m=M\quad(\mathrm{tag}), \\  \rule[-2mm]{0mm}{5mm}
\delta_{m'm} -\delta_{m',m-1} &:~ m=M\quad(\mathrm{cap}).
\end{array} \right.
\end{equation}

The exact statistical mechanical analysis (\ref{eq:5})-(\ref{eq:7}) yields the (scaled) Gibbs free energy,
\begin{equation}\label{eq:a3} 
\beta\bar{G}\doteq-\lim_{N\to\infty}N^{-1}\ln Z
=-\ln\Big(1+w_1^{-1}\Big),
\end{equation}
where 
\begin{equation}\label{eq:a4} 
w_m=\left\{\begin{array}{ll} \rule[-2mm]{0mm}{5mm}
e^{\beta\epsilon_m}-1 &:~ m=M\quad (\mathrm{tag}), \\ \rule[-2mm]{0mm}{5mm}
e^{\beta\epsilon_m} &:~ m=M\quad (\mathrm{cap}), \\ \rule[-2mm]{0mm}{5mm}
{\displaystyle e^{\beta\epsilon_m}\frac{w_{m+1}}{1+w_{m+1}}} &:~ m=M-1,\ldots,1
\end{array} \right.
\end{equation}
are determined recursively. 
For the population densities of nested extension particles we obtain
\begin{equation}\label{eq:a5} 
\bar{N}_m=\left\{\begin{array}{ll} \rule[-2mm]{0mm}{5mm}
{\displaystyle \prod_{m'=1}^{m}\frac{1}{1+w_{m'}}} &:~ m=1,\ldots,M-1, \\ \rule[-2mm]{0mm}{6mm}
\bar{N}_{M-1}/w_M &:~ m=M\quad (\mathrm{tag}), \\ \rule[-2mm]{0mm}{5mm}
\bar{N}_{M-1}/(1+w_M) &:~ m=M\quad (\mathrm{cap}).
\end{array} \right.
\end{equation}

%
\section{Enthalpic elasticity}\label{sec:for-ext}
%
Consider an open polymeric chain of $N$ monomers under controllable tension.
The reference state in our modeling [Fig.~\ref{fig:figure1}(a)], most closely realized at low tension, has the chain straightened out to its contour length without significant contour elongation.
Populating this pseudovacuum with extension particles via thermal or mechanical activation describes the phenomenon of contour elasticity, which is predominantly enthalpic in nature.
Conversely, the phenomenon of thermal unbending toward the same reference state is a consequence of the deactivation of contraction particles (with negative excess length), which is the topic of Sec.~\ref{sec:appe}. Thermal unbending is predominantly entropic in nature.
In the following we discuss aspects of increasing complexity related to contour elasticity. 

\subsection{One-step elasticity}\label{sec:1-ste-ela}
\begin{figure}[b]
  \begin{center}
\includegraphics[width=40mm]{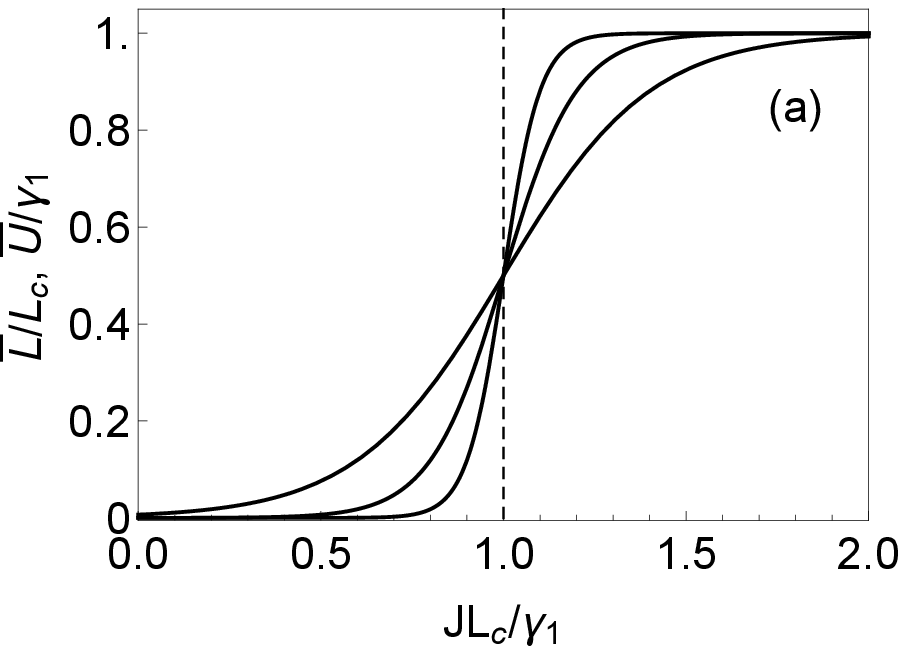}\hspace*{3mm}\includegraphics[width=40mm]{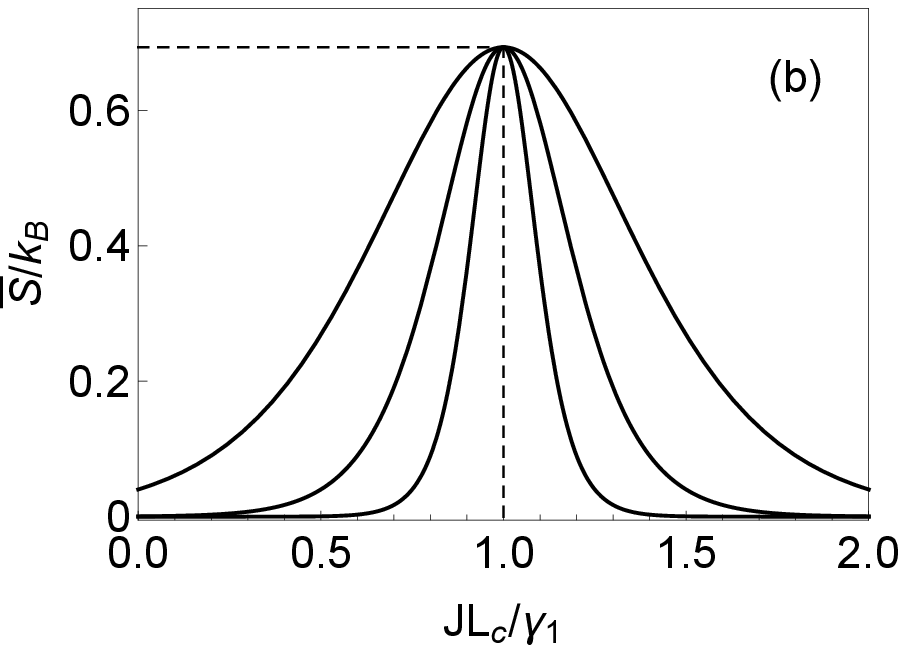}
\includegraphics[width=40mm]{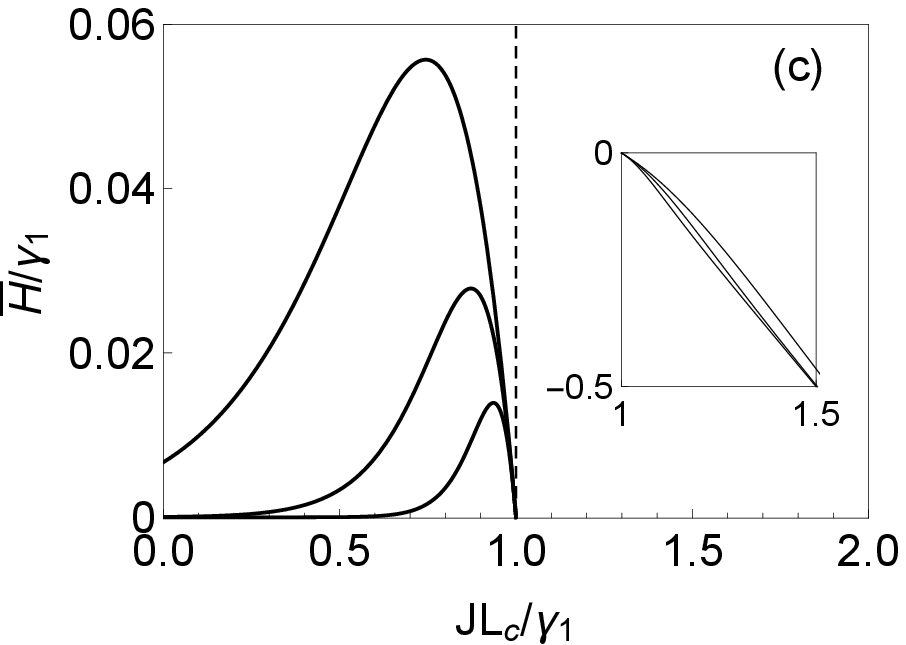}\hspace*{3mm}\includegraphics[width=40mm]{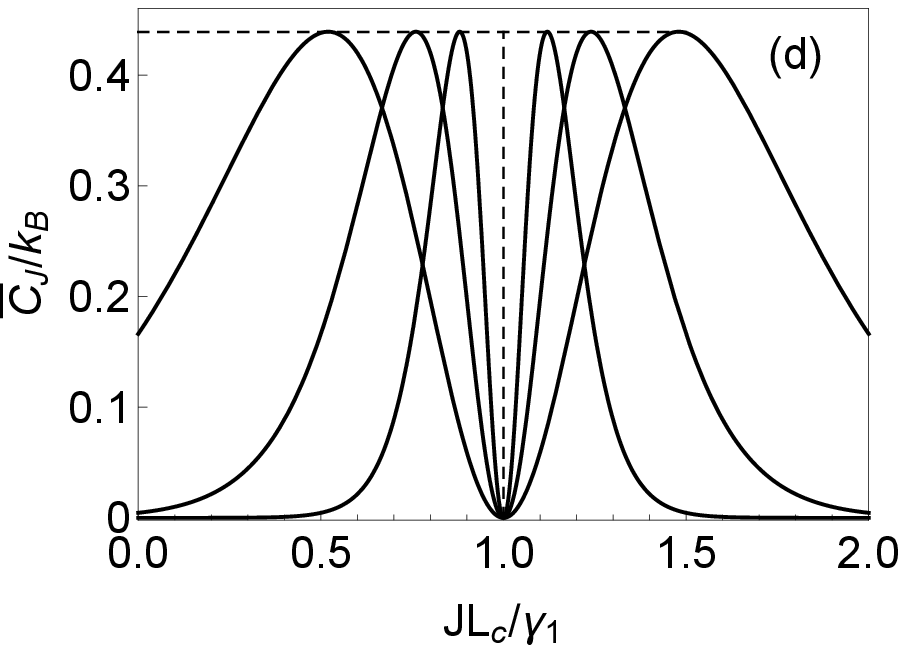}
\includegraphics[width=40mm]{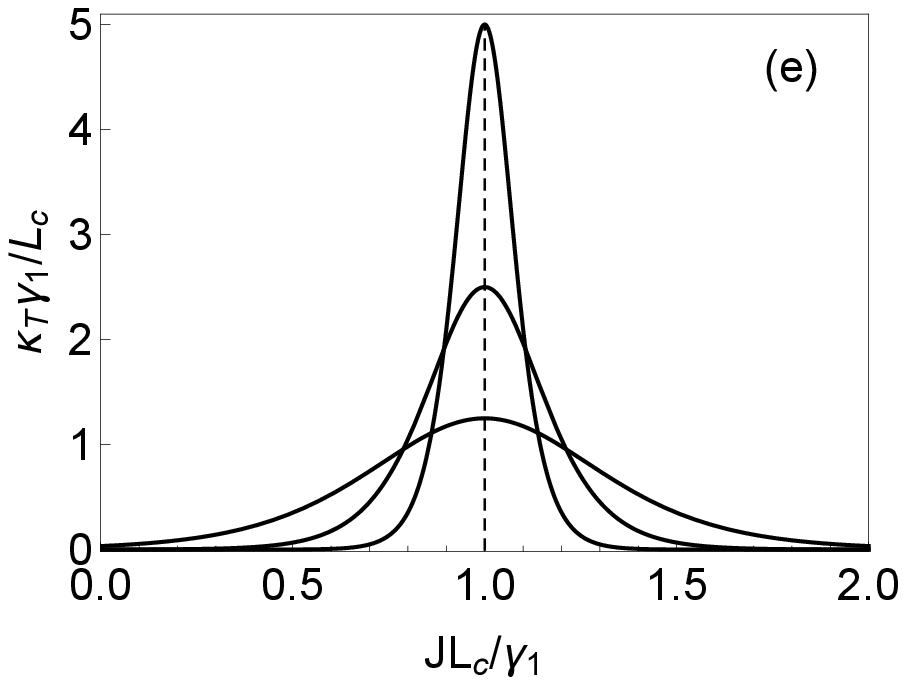}\hspace*{3mm}\includegraphics[width=40mm]{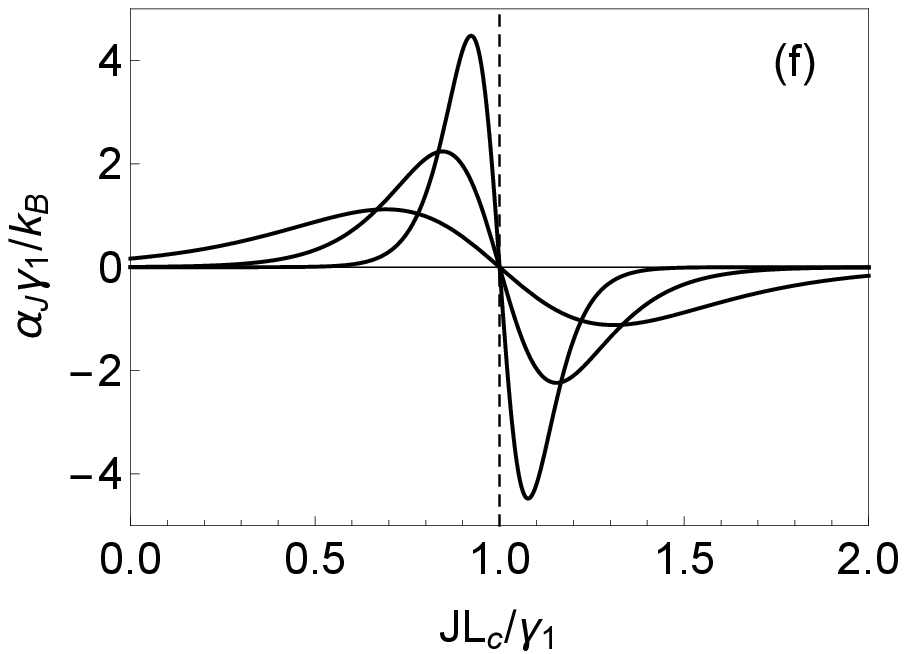}
\end{center}
\caption{(a) Scaled excess length and internal energy, (b) entropy, (c) enthalpy, (d) heat capacity, (e) tensile compliance, and (f) thermal expansivity, all versus scaled tension at constant scaled temperature $k_\mathrm{B}T/\gamma_1=0.05, 0.1, 0.2$. The horizontal dashed lines mark the $T$-independent peak values, $\bar{S}_\mathrm{max}/k_\mathrm{B}=\ln2=0.693\ldots$ and $\bar{C}_\mathrm{max}=0.439\ldots$.}
  \label{fig:figure2}
\end{figure}

Polymeric bonds described by a single species of compact level-1 extension particles $(M=1)$ are elastic in a very primitive sense.
Each bond is either fully extended to $L_1=L_\mathrm{c}$ or not extended at all.
The sum $B_{00}$ in the partition function (\ref{eq:45}) has two terms only.
The Gibbs free energy (\ref{eq:13}) becomes
\begin{equation}\label{eq:22} 
\bar{G}=-k_\mathrm{B}T\ln\Big(1+e^{-K_1}\Big),\quad K_1=\beta(\gamma_1-JL_\mathrm{c}).
\end{equation}
The extension particles have the statistics of lattice fermions, reflected in the average excess length,
\begin{equation}\label{eq:23} 
\bar{L}=L_1\bar{N}_1=\frac{L_\mathrm{c}}{e^{K_1}+1},
\end{equation}
and also recognizable in the entropy expression derived from (\ref{eq:8}) or (\ref{eq:15}),
\begin{align}\label{eq:24} 
\frac{\bar{S}}{k_\mathrm{B}} &=-\bar{N}_1\ln\bar{N}_1-(1-\bar{N}_1)\ln(1-\bar{N}_1) \nonumber \\
&=\ln\Big(1+e^{-K_1}\Big)+\frac{K_1}{e^{K_1}+1}.
\end{align}
The enthalpy, the internal energy, and the three response functions become
\begin{equation}\label{eq:25} 
\bar{H}=\epsilon_1\bar{N}_1=\frac{\gamma_1-JL_\mathrm{c}}{e^{K_1}+1},
\end{equation}
\begin{equation}\label{eq:26} 
\bar{U}=\frac{\gamma_1}{e^{K_1}+1},\quad 
\bar{C}_J=k_\mathrm{B}\frac{K_1^2}{4\cosh^2(K_1/2)},
\end{equation}
\begin{equation}\label{eq:27} 
\kappa_T=\frac{\beta L_\mathrm{c}}{4\cosh^2(K_1/2)},\quad \alpha_J=\frac{K_1/T}{4\cosh^2(K_1/2)}.
\end{equation}
These quantities as functions of tension at constant temperature are shown in Fig.~\ref{fig:figure2}.

Only at $T=0$ does the one-step elasticity produce a step in the force-extension characteristic [Fig.~\ref{fig:figure2}(a)].
With $T$ increasing from zero the step elasticity of bonds is subject to thermal fluctuations and produces force-extension curves whose rise from zero takes place over a widening interval of applied tension.
The same characteristic with different scaling holds for the internal energy. 
Every bond that is extended contains the same unit of elastic energy.
That energy is supplied as extension work or as heat from the reservoir.
The thermal activation of bond extensions is most effective at tensions near $JL_\mathrm{c}/\gamma_1=1$, when the activation energy $\epsilon_1$ is near zero. 

The configurational entropy [Fig.~\ref{fig:figure2}(b)] is associated with the random distribution of extended bonds.
The entropy curves start low at zero tension.
Only few bonds are extended because the particle activation energy is high.
The curves go through a maximum when half the bonds are extended (on average) and return back to low values at high tension as the number of non-extended bonds shrinks to near zero.
Here the activation energy of extended bonds is negative and large in magnitude compared to $k_\mathrm{B}T$.

The average activation energy per monomer is reflected in the enthalpy curves [Fig.~\ref{fig:figure2}(c)].
With the tension gradually increasing from zero, this measure goes up even though the activation energy of individual particles goes down. 
However, the enthalpy curves turn sharply toward negative values as the activation energy changes sign.
To the right of the dashed line, the number of activated particles is near saturation.
Here the enthalpy curves reflect the (linear) dependence of the particle activation energy on the tension.

The rise and fall of the entropy with increasing tension is associated with heat transfer between the chain and the surrounding fluid.
The chain absorbs heat while its macrostate becomes more disordered and then expels heat when further stretching restores ordering. 
Figure~\ref{fig:figure2}(d) shows that the two regions of maximum heat transfer (forced mechanically via change in tension) coincide with regions of maximum heat capacity at constant tension.

The two mechanical response functions exhibit contrasting features.
The tensile compliance [Fig.~\ref{fig:figure2}(e)] reflects the slope of the force-extension characteristic.
It has a maximum at $JL_\mathrm{c}/\gamma_1=1$.
That maximum becomes higher and narrower with decreasing temperature.
The thermal expansivity [Fig.~\ref{fig:figure2}(f)], by contrast, switches sign at $JL_\mathrm{c}/\gamma_1=1$. 
Below that value of tension, the chain expands when it is heated up whereas above that value the chain expands when it is cooled down. 
At low $T$ this effect is more pronounced than at high $T$ and it takes place within a narrower range of tension.

\subsection{Cooperativity}\label{sec:coop}
The model of one-step elasticity can be adapted to describe a structural conversion between two conformations of the molecular chain under controlled tension.
For that purpose, cooperativity effects must be built in.
The native conformation, energetically favored at low tension, is the reference state.
With increasing tension, segments of an extended conformation are being nucleated out of this state.
The segments of extended conformation then grow at the expense of segments in the native conformation as the tension continues to increase.

The distribution of segment lengths at given tension depends on the difference between the energies required to nucleate extended segments and to grow such segments.
If the former is larger than the latter cooperativity ensues.  
The average number of segments is suppressed and the average length is enhanced relative to a random distribution of interfacial points. 

We can add controllable cooperativity to our model of one-step elasticity by switching from a single species of particles $(M=1)$ at level 1 (Sec.~\ref{sec:1-ste-ela}) to $M=2$ species of nested particles at level 2 [Fig.~\ref{fig:figure6}(e)], namely one host $(m=1)$ and one tag $(m=2)$.
The combinatorial specifications are
\begin{equation}\label{eq:46} 
 A_1=N-2,\quad A_2=0,\quad
 \mathbf{g}=\left(
 \begin{array}{rr} 2 & ~1 \\ -1 & ~0 \end{array} \right).
\end{equation} 
The statistical interaction coefficients $g_{11}=2$, $g_{12}=1$ mean that placing a host (tag) removes two slots (one slot) available to place further hosts.
The coefficient $g_{21}=-1$ tells us that by placing a host, which nucleates a segment of extended conformation, we open up a new slot for placing tags.
Adding a tag, which grows an existing segments, leaves the number of slots for the placement of further tags invariant. Hence $g_{22}=0$.

The activation energies (\ref{eq:9}) are used here with parameter settings,
\begin{equation}\label{eq:47} 
L_1=L_2\doteq L_\mathrm{c},\quad \gamma_1\geq\gamma_2>0.
\end{equation}
Equations~(\ref{eq:6}) then reduce to a quadratic equation.
The physically relevant solution, expressed as a function of the two parameters,
\begin{equation}\label{eq:28} 
K_i=\beta(\gamma_i-JL_\mathrm{c})\quad :~ i=1,2,
\end{equation}
reads
\begin{align}\label{eq:29} 
& w_1=w_2e^{K_1-K_2}=
e^{K_1-K_2/2}\big[\sinh(K_2/2)+\xi(K_1,K_2)\big], \nonumber \\
&\xi(K_1,K_2)=\sqrt{\sinh^2(K_2/2)+e^{K_2-K_1}}.
\end{align}
The growth parameter $K_2$ controls the total length of segments in the extended conformation under variable tension, whereas the nucleation parameter $K_1-K_2$ controls the number of such segments.
At low cooperativity $(K_1\gtrsim K_2)$ the segments of extended conformation tend to be short and numerous.
At high cooperativity $(K_1\gg K_2)$ they are longer and fewer in number on average \cite{note1}.

\begin{figure}[htb]
  \begin{center}
\includegraphics[width=40mm]{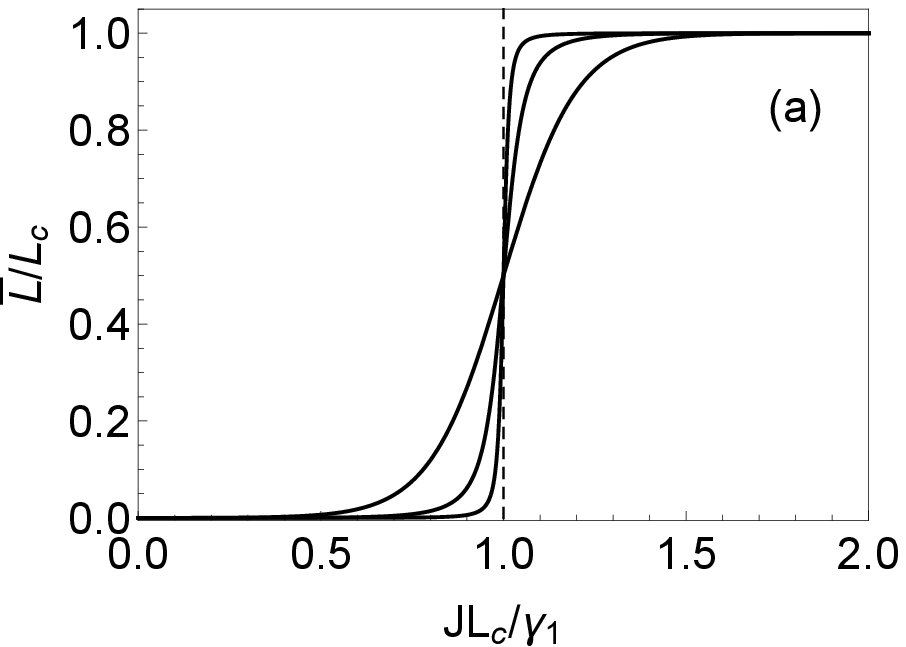}\hspace*{3mm}\includegraphics[width=40mm]{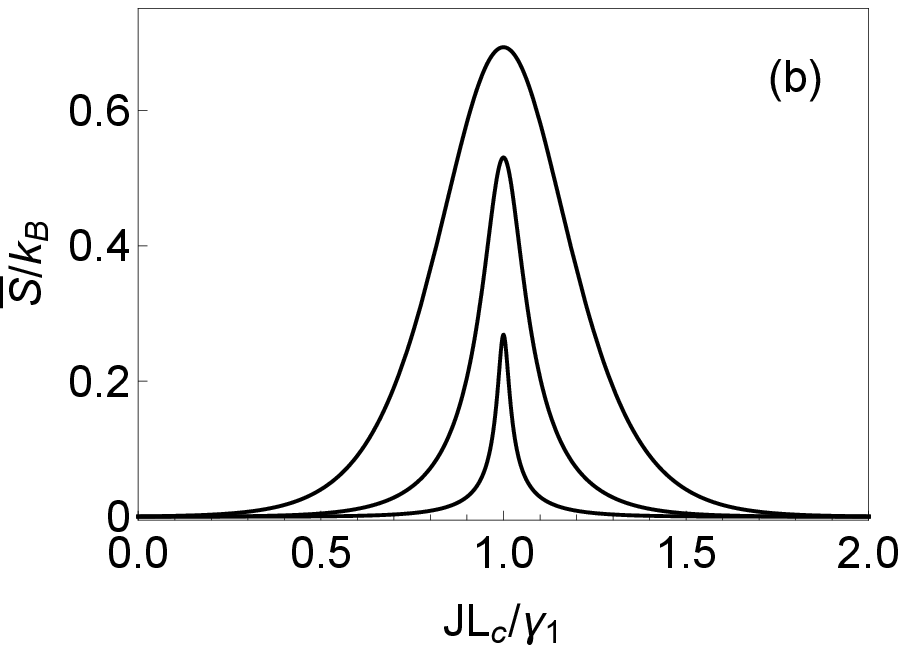}
\includegraphics[width=40mm]{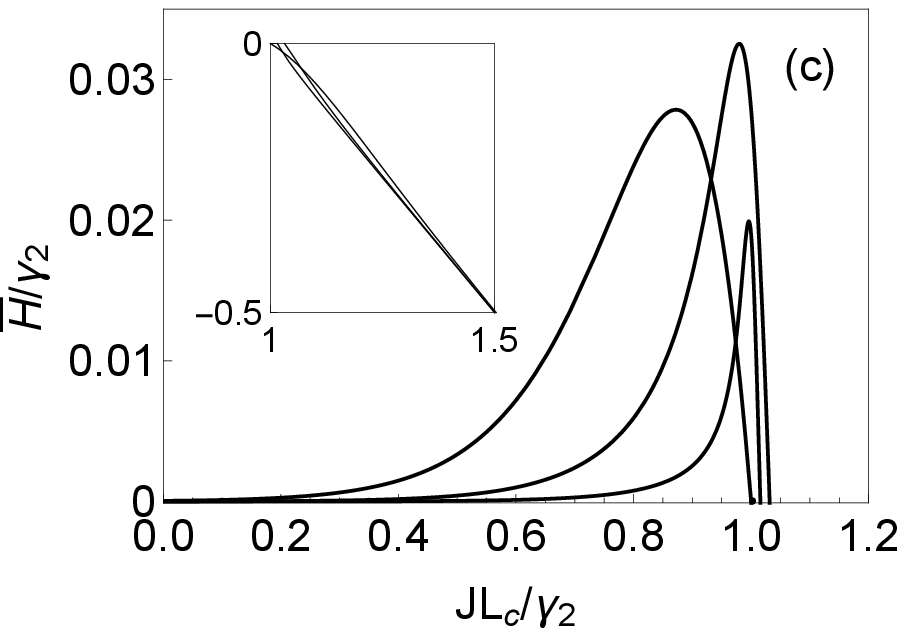}\hspace*{3mm}\includegraphics[width=40mm]{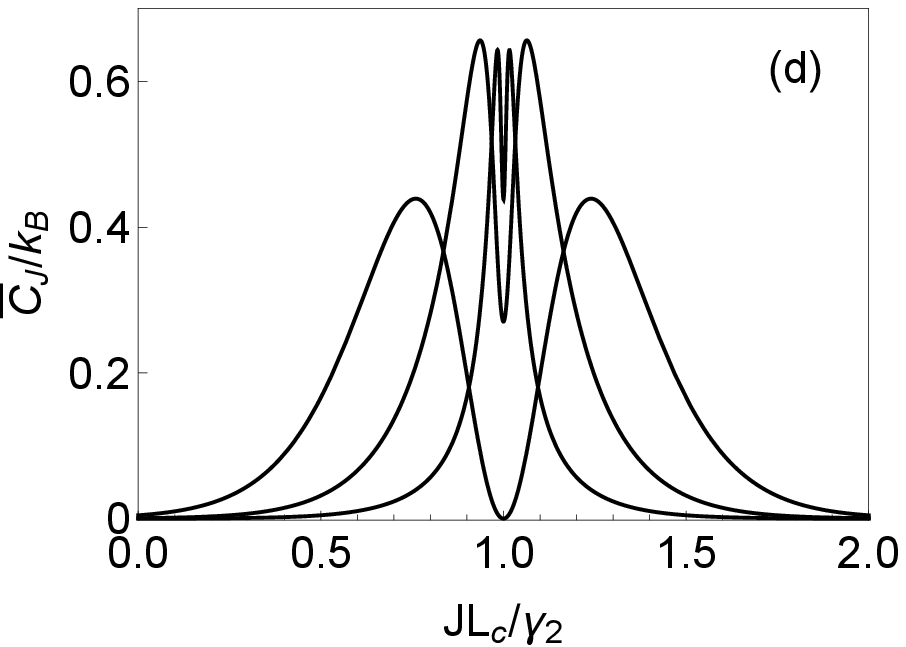}
\end{center}
\caption{(a) Scaled excess length, (b) entropy, (c) enthalpy, and (d) heat capacity, all versus scaled tension at constant scaled temperature $k_\mathrm{B}T/\gamma_2=0.1$ and cooperativity $K_1-K_2=0, 0.2, 5.0$ (broad to sharp features).}
  \label{fig:figure14}
\end{figure}

The Gibbs free energy inferred from the partition function (\ref{eq:5}) with $w_1$ from (\ref{eq:29}) becomes
\begin{align}\label{eq:30} 
\beta\bar{G} &=-\ln\big(1+w_1^{-1}\big) \nonumber \\
&= K_2/2-\ln\big(\cosh(K_2/2)+\xi(K_1,K_2)\big).
\end{align}
Quantities of interest are then inferred via (\ref{eq:14})-(\ref{eq:19}).
More or less compact expressions which extend the results (\ref{eq:23})-(\ref{eq:27}) to account for effects of cooperativity include
\begin{align}\label{eq:72}
  \bar{L} = \frac{L_\mathrm{c}}{2}
  \left[1-\frac{\sinh (K_{2}/2)}{\xi(K_{1},K_{2})}\right],
\end{align}
\begin{align}\label{eq:73}
  \frac{\bar{S}}{k_{\mathrm{B}}}
  &=
    \ln\big(\cosh(K_{2}/2)+\xi(K_{1},K_{2})\big)
    -\frac{K_{2}}{2}\frac{\sinh(K_{2}/2))}{\xi(K_{1},K_{2})} \nonumber \\
  & \quad +\frac{K_{1}-K_{2}}{2}
    \left(1-\frac{\cosh(K_{2}/2)}{\xi(K_{1},K_{2})}\right),
\end{align}
\begin{align}\label{eq:74}
  \beta\bar{H} &= \frac{K_{1}}{2}
   -\frac{1}{2\xi(K_{1},K_{2})}
  \Big[K_{2}\sinh\frac{K_{2}}{2} \nonumber \\
  &\hspace{30mm}+(K_{1}-K_{2})\cosh\frac{K_{2}}{2}\Big],
\end{align}
for excess length, entropy, and enthalpy, respectively.
The graphs in Fig.~\ref{fig:figure14} are designed to illustrate effects of cooperativity in four quantities.
Focusing on the intermediate temperature selected in Fig.~\ref{fig:figure2}, we strengthen the cooperativity in two steps.

The effect on the excess length is similar to that of decreasing temperature.
The same is true with only very small differences in the scaled curves for the internal energy (not shown). 
Decreasing temperature or increasing cooperativity both suppress entropy but in  different ways. In both cases the entropy peak becomes narrower but only cooperativity makes it also shorter.
Mixed segments produce less entropy when their sizes increase and their number decreases.
The enthalpy peak becomes smaller with decreasing temperature and sharper with increasing cooperativity.
The heat capacity peaks keep the same height under varying temperature but become taller with increasing cooperativity. 
In the other two response functions (not shown) the effects are more similar.

\subsection{Rupture}\label{sec:rupt}
A system of compact level-1 extension particles with an infinite number of species $(M=\infty)$ and energetic specifications, 
\begin{equation}\label{eq:31} 
L_m=mL_\mathrm{c},\quad \gamma_m=m\gamma_\mathrm{c}\quad :~ m=1,2,\ldots,
\end{equation} 
is set up for a catastrophic event when the tension reaches a critical value, $J_\mathrm{crit}=\gamma_\mathrm{c}/L_\mathrm{c}$, where all activation energies go negative.
Further extension requires no work.
The polymeric chain suffers a rupture.

\begin{figure}[htb]
  \begin{center}
\includegraphics[width=40mm]{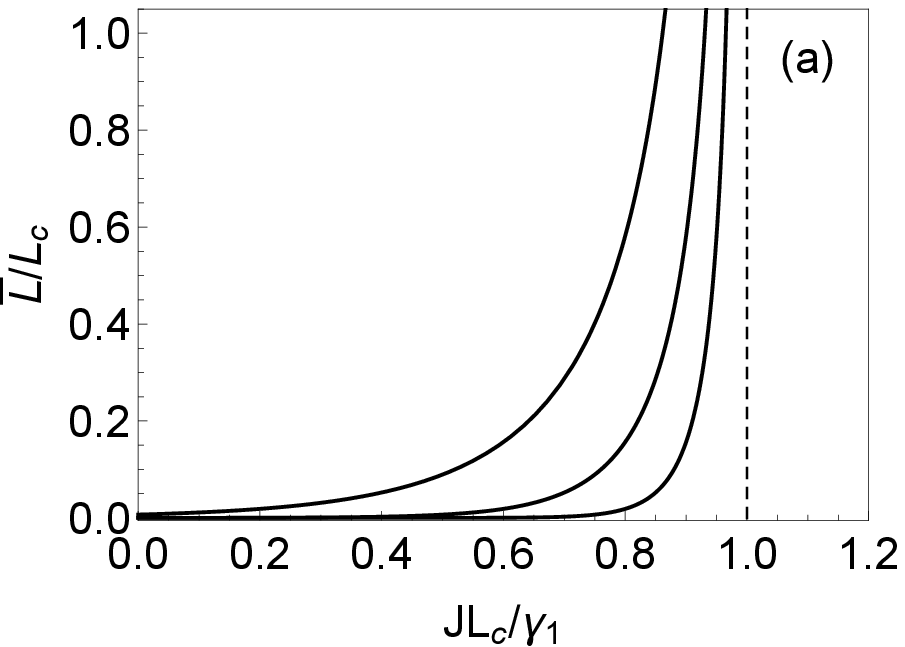}\hspace*{3mm}\includegraphics[width=40mm]{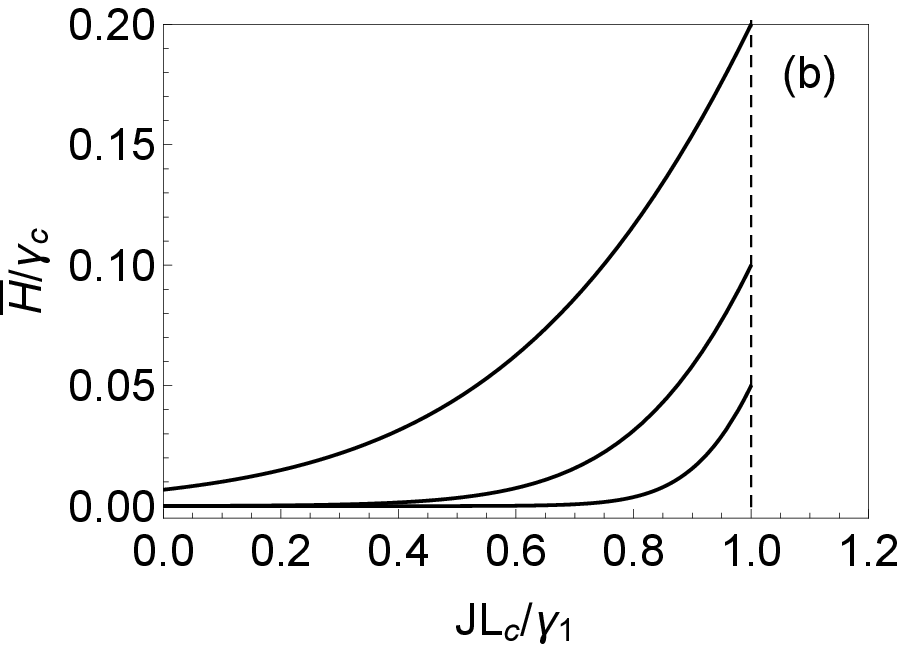}
\end{center}
\caption{(a) Scaled excess length and (b) enthalpy versus scaled tension at constant scaled temperature $k_\mathrm{B}T/\gamma_1=0.05, 0.1, 0.2$.}
  \label{fig:figure3}
\end{figure}

The sum $B_{00}$ in (\ref{eq:12}) becomes a geometric series and yields
the result,
\begin{equation}\label{eq:32} 
\bar{G}=k_\mathrm{B}T\ln\Big(1-e^{-K_\mathrm{c}}\Big),\quad K_\mathrm{c}=\beta(\gamma_\mathrm{c}-JL_\mathrm{c}),
\end{equation}
for the Gibbs free energy (\ref{eq:13}).
The excess length and the enthalpy inferred from (\ref{eq:14}) and (\ref{eq:16}),
\begin{equation}\label{eq:33} 
\bar{L}=\frac{L_\mathrm{c}}{e^{K_\mathrm{c}}-1},\quad 
\bar{H}=\frac{\gamma_1-JL_\mathrm{c}}{e^{K_\mathrm{c}}-1},
\end{equation}
encounter singularities at $J_\mathrm{crit}$ indicative of rupture as illustrated in Fig.~\ref{fig:figure3}.
The average excess length diverges whereas the enthalpy curves terminate in cusps at the critical tension.
As the particles proliferate catastrophically, their activation energies approach zero while their contributions to extra length remain undiminished

There are singularities at $J_\mathrm{crit}$ in other quantities as well. The results for the entropy, internal energy, and response functions read
\begin{align}\label{eq:34} 
\frac{\bar{S}}{k_\mathrm{B}} =\frac{K_\mathrm{c}}{e^{K_\mathrm{c}}-1}-\ln\Big(1-e^{-K_\mathrm{c}}\Big),
\end{align}
\begin{equation}\label{eq:35} 
\bar{U}=\frac{\gamma_\mathrm{c}}{e^{K_\mathrm{c}}-1},\quad 
\bar{C}_J=k_\mathrm{B}\frac{K_\mathrm{c}^2}{4\sinh^2(K_\mathrm{c}/2)},
\end{equation}
\begin{equation}\label{eq:36} 
\kappa_T=\frac{\beta L_\mathrm{c}}{4\sinh^2(K_\mathrm{c}/2)},\quad 
\alpha_J=\frac{K_\mathrm{c}/T}{4\sinh^2(K_\mathrm{c}/2)}.
\end{equation}
All these quantities diverge at $J_\mathrm{crit}$.
The rupture announces itself, as $J$ approaches $J_\mathrm{crit}$ from below, by a heightened sensitivity to changes in both tension and temperature. 

We have to keep in mind that rupture is a dynamic phenomenon whereas our methodology describes equilibrium states. 
The latter can only describe certain threshold aspects of the former.
For example, the divergent $\bar{U}$, which represents an infinite elastic energy at $J_\mathrm{crit}$, would only be real if the tension were maintained at the critical value as the excess length diverges.
In a dynamic rupture event that is not the case.
The tension collapses before the excess length diverges.

The size distribution of compact extension particles is geometric in nature (Pascal distribution):
\begin{equation}\label{eq:37} 
\bar{N}_m=e^{-mK_\mathrm{c}}\big(1-e^{-K_\mathrm{c}}\big)\quad :~ m=1,2,\ldots
\end{equation}
The smallest sizes dominate at low tension.
The distribution becomes broad and flat as $J$ approaches $J_\mathrm{crit}$.
Each size-$m$ particle effectively represents $m$ bosons.
The number of bosons in the distribution (\ref{eq:37}) of compacts then becomes 
\begin{equation}\label{eq:38} 
\bar{N}_\mathrm{bos}\doteq\sum_{m=1}^\infty m\bar{N}_m=\frac{1}{e^{K_\mathrm{c}}-1}
\end{equation}
and the entropy (\ref{eq:34}) can be rendered in the (recognizably bosonic) form
\begin{equation}\label{eq:39} 
\frac{\bar{S}}{k_\mathrm{B}} =(1+\bar{N}_\mathrm{bos})\ln(1+\bar{N}_\mathrm{bos})-\bar{N}_\mathrm{bos}\ln \bar{N}_\mathrm{bos}.
\end{equation}

We can replace infinitely many species of compacts by just two species of nested particles: hosts and tags from Sec.~\ref{sec:nes-ext-par-1} with specifications,
\begin{equation}\label{eq:48}
L_\mathrm{h}=L_\mathrm{t}=L_\mathrm{c},\quad \gamma_\mathrm{h}=\gamma_\mathrm{t}=\gamma_\mathrm{c}.
\end{equation}
The results (\ref{eq:32})-(\ref{eq:36}) are readily reproduced.
The population densities of hosts and tags add up to the boson distribution (\ref{eq:38}):
\begin{equation}\label{eq:49} 
\bar{N}_\mathrm{h}+\bar{N}_\mathrm{t}
=e^{-K_\mathrm{c}}+\frac{e^{-K_\mathrm{c}}}{e^{K_\mathrm{c}}-1}
=\frac{1}{e^{K_\mathrm{c}}-1}.
\end{equation}

\subsection{Linear contour elasticity}\label{sec:hooke}
\begin{figure}[htb]
  \begin{center}
\includegraphics[width=41mm]{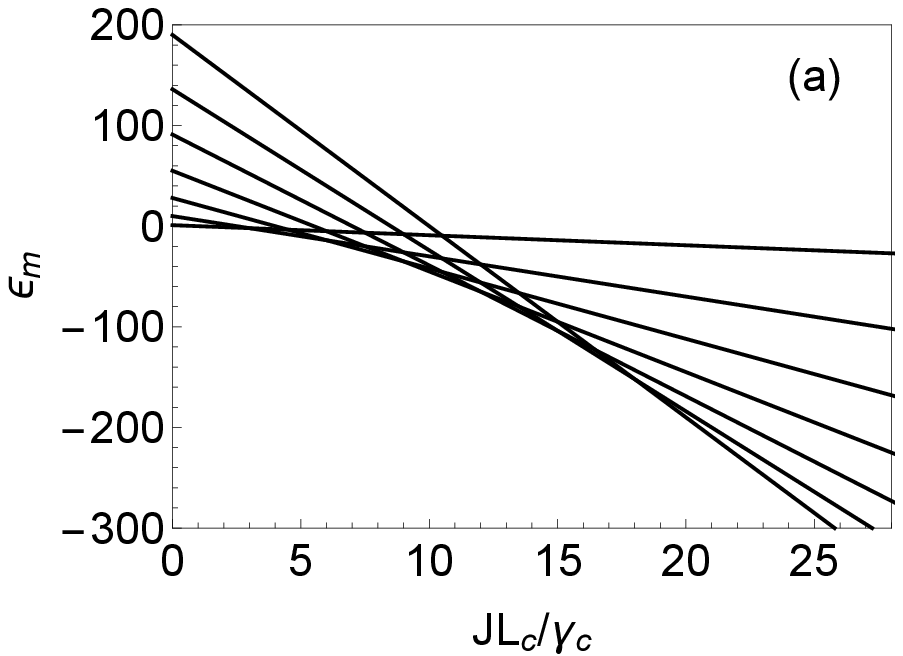}\hspace*{3mm}\includegraphics[width=39mm]{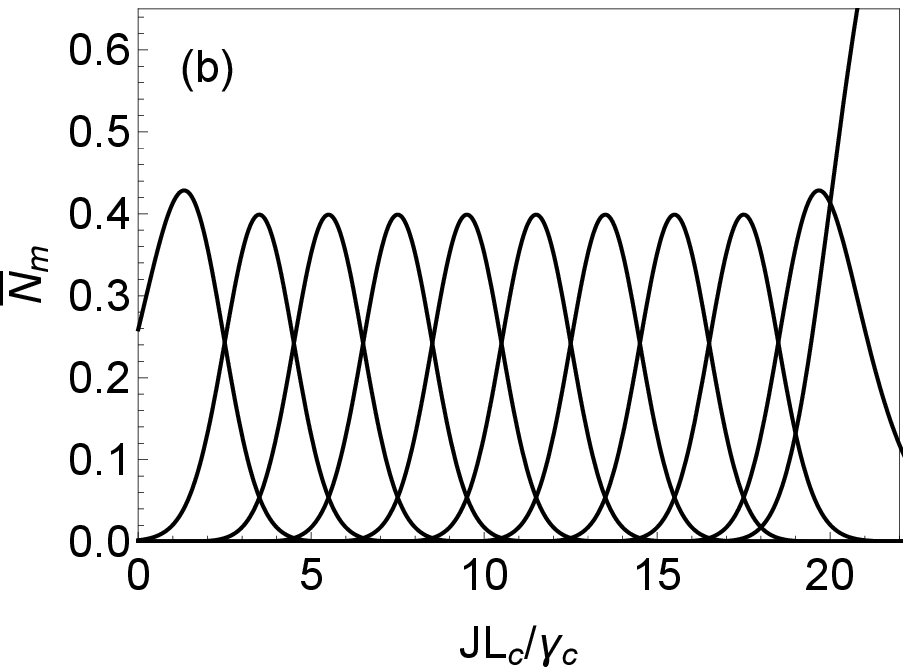}
\end{center}
\caption{(a) Particle activation energies $\epsilon_m$, $m=1$, 4, 7, $\ldots$, 19 (from top down on the right), and (b) particle population densities $\bar{N}_m$, $m=1,3,5,\ldots,19,20$, (from left to right) versus scaled tension. The constant scaled temperature used is $k_\mathrm{B}T/\gamma_\mathrm{c}=1$.}
  \label{fig:figure5}
\end{figure}

From compact level-1 extension particles we now construct a model of linear elasticity.
We choose activation energies (\ref{eq:9}) with specifications,
\begin{equation}\label{eq:40} 
L_m=mL_\mathrm{c},~ \gamma_m=\frac{1}{2}m(m+1)\gamma_\mathrm{c}~:~ m=1,\ldots,M.
\end{equation}
The (negative) slopes of their (linear) $J$-dependence vary linearly in $m$ and the intercepts quadratically [Fig.~\ref{fig:figure5}(a)].
As $J$ increases from zero, $\epsilon_1$ goes negative first, then $\epsilon_2$ becomes more negative than $\epsilon_1$, next $\epsilon_3$ overtakes $\epsilon_2$ etc.  The crossing between $\epsilon_m$ and $\epsilon_{m-1}$ takes place at $J L_\mathrm{c}=m\gamma_\mathrm{c}$. 
At this tension the excess length $\bar{L}$ of the lowest-energy state jumps from $(m-1)L_\mathrm{c}$ to $mL_\mathrm{c}$. 

At nonzero but not too high temperatures, only the lowest few levels are populated with significant probability [Fig.~\ref{fig:figure5}(b)].
As $J$ increases, probability is gradually shifted from compacts with smaller excess length to compacts with larger excess length.
The latter crowd out the former.
The probability distributions have nearly the same shape for all compacts except near the beginning or the end of the sequence.
These attributes produce Hookean elasticity. 

\begin{figure}[b!]
  \begin{center}
\includegraphics[width=40mm]{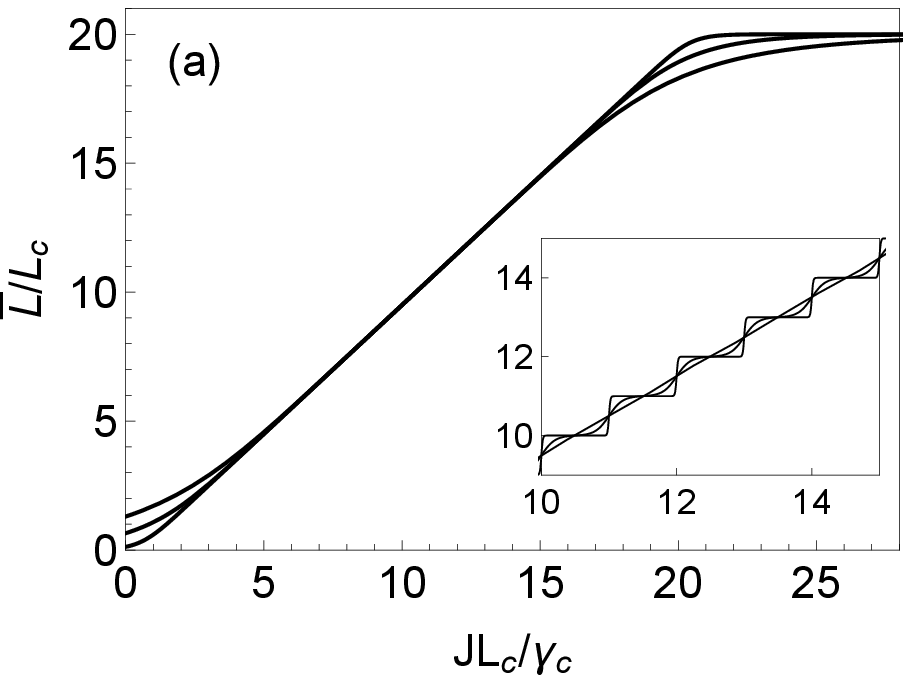}\hspace*{3mm}\includegraphics[width=40mm]{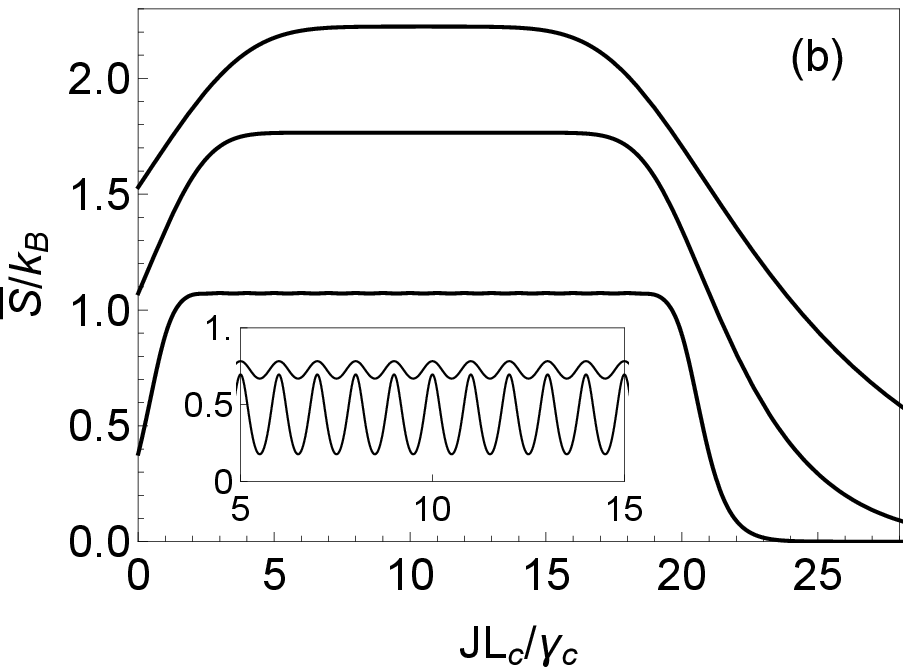}
\includegraphics[width=40mm]{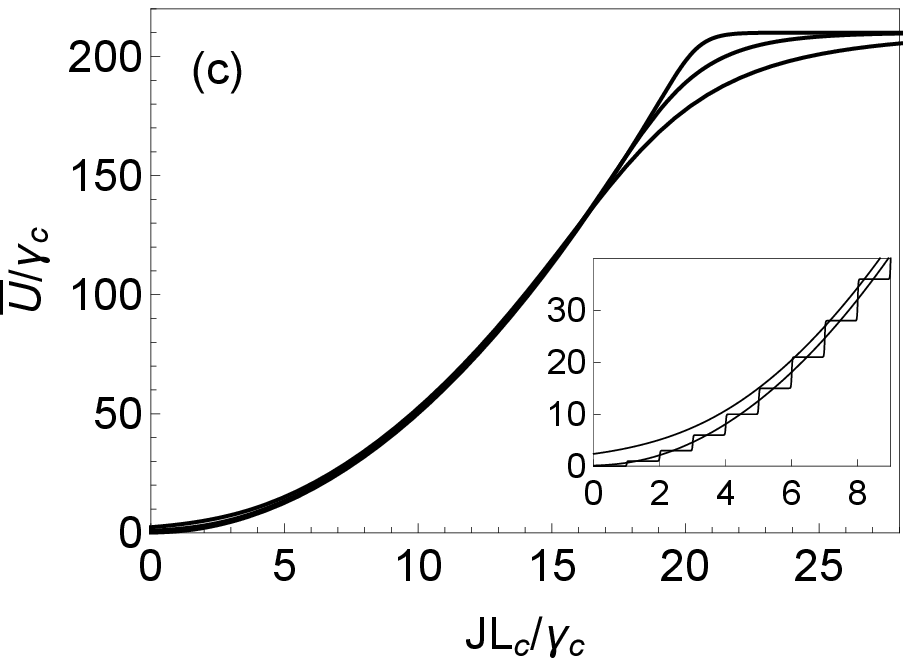}
\end{center}
\caption{(a) Scaled excess length, (b) entropy, and (c) internal energy, all versus scaled tension at constant scaled temperature $k_\mathrm{B}T/\gamma_\mathrm{c}=0.5, 2, 5$ for the elastic model with specifications (\ref{eq:40}) and $M=20$. 
In the insets we show data for $k_\mathrm{B}T/\gamma_\mathrm{c}=0.25,0.1,0.01$, $k_\mathrm{B}T/\gamma_\mathrm{c}=0.25, 0.125$, and $k_\mathrm{B}T/\gamma_\mathrm{c}=5,0.5,0.01$, respectively.}
  \label{fig:figure4}
\end{figure}

Data for quantities of interest are inferred from expressions (\ref{eq:13})-(\ref{eq:19}). 
A selection is shown in Figs.~\ref{fig:figure4}(a)-(c).
Over a considerable range of temperature, the force-extension characteristic is linear, representing the familiar elasticity of a rubber band [Fig.~\ref{fig:figure4}(a)]. 
Deviations are largest at low tension and high tension.
Real elastic materials obey Hooke's law only for tensions of a limited range. 
At the low end of that range, enthalpic elasticity crosses over into entropic elasticity, which is the topic of Sec.~\ref{sec:appe}.
At the high end, many materials turn stiff and resist further stretching until they rip.
That limiting stiffness is naturally accounted for in our model by the finite number $M$ of species of extension particles.

At low $J$ all activation energies are positive and at high $J$ all are negative.
Raising $T$ has opposite effects at low and high tension. It thermally activates extension particles at low $J$ and thermally annihilates them at high $J$.
This leads to expansion in the first instance and to contraction in the second.

At low $T$, the discrete nature of our model becomes more conspicuous unless we increase $M$.
The linear elasticity as represented in the force-extension characteristic degenerates into a staircase of uniform steps at uniform intervals as shown in the inset to Fig.~\ref{fig:figure4}(a). 
The internal energy varies quadratically with tension, again over a significant range of tension in true Hookean fashion and for a significant range of temperatures [Fig.~\ref{fig:figure4}(c)].
The discrete quanta of elastic energy become visible at sufficiently low $T$ as shown in the inset.

The variation of entropy with tension and temperature [Fig.~\ref{fig:figure4}(b)] is more complex.
The most conspicuous feature is the plateau over a range of $J$ at a level that increases with $T$.
Such behavior is expected for a molecular chain of bonds responding linearly to tension.
The spectrum of low-lying excitations remains harmonic-oscillator-like, meaning that the level-spacings are independent of $J$.
We note that the $J$-independent stretch of $\bar{S}$ is a consequence of the shape of the particle population distributions $\bar{N}_m$ described earlier [Fig.~\ref{fig:figure5}(b)].
Raising $T$ at constant $J$ broadens them, causing an increase in $\bar{S}$. 

At low $T$, specifically at $k_\mathrm{B}T/\gamma_\mathrm{c}\lesssim 0.25$, the structure of the entropy curve is qualitatively different. 
Right above that threshold, the curve has a very flat portion at height $\bar{S}/k_\mathrm{B}\gtrsim\ln2\simeq 0.69$.
All extension particles are frozen out except the two with the lowest activation energies. 
With the tension increasing, the low lying states change as in a relay race but it is always lowest two that have significant occupancy.
The entropy of mixing remains constant.
Below that threshold, the lowest one has a significantly higher occupancy than the second lowest except near the crossing point.
Hence the entropy oscillates between a low value that is strongly $T$-dependent and a high value at $\bar{S}/k_\mathrm{B}=\ln2$.
An increase in $M$ combined with some rescaling will suppress the low-$T$ structures highlighted in all insets to Fig.~\ref{fig:figure4}.

This model of linear elasticity can be transcribed from $M$ compacts with specifications (\ref{eq:40}) to an equivalent system of $M$ level-1 nested particles: one host, $M-2$ hybrids, and one cap from Sec.~\ref{sec:nes-ext-par-1} with specifications,
\begin{equation}\label{eq:50}
L_m=L_\mathrm{c},\quad \gamma_m=m\gamma_\mathrm{c}\quad :~ m=1,\ldots,M.
\end{equation}

In the application to DNA (Sec.~\ref{sec:DNA-1}), the regime of contour elasticity is preceded by a regime of thermal unbending at lower tension (Sec.~\ref{sec:appe}). 
This requires some adaptation in the design of activation energies for level-1 compact extension particles (Appendix~\ref{sec:appa}).

\subsection{Power-law contour elasticity}\label{sec:appf}
Nonlinear responses of thermodynamic systems to forces exerted by external agents are quite common, e.g. near critical points.
Here we analyze a scenario of power-law elastic response to tension by a mere change in the specifications of the level-1 compact extension particles used above.

Our methodology demands that any $J$-dependence of particle activation energy is linear.
This criterion is satisfied by the activation energies (\ref{eq:9}) with specifications (\ref{eq:40}) or (\ref{eq:50}), producing a linear (Hookean) force-extension characteristic.
We can generalize the force-extension characteristic to a power-law behavior, $\bar{L}\sim J^\nu$, $0<\nu<2$, with a set, $m=1,\ldots,M$, of level-1 compacts provided we generalize the specification of their activation energies (\ref{eq:9}) from (\ref{eq:40}) to
\begin{equation}\label{eq:f1} 
L_m=m^\nu L_\mathrm{c},\quad  
\gamma_m=\gamma_\mathrm{c}\sum_{n=1}^mn\left[(n+1)^\nu-n^\nu\right].
\end{equation}
In all results and graphs that follow we define the scales for energy, length, and tension by setting $\gamma_\mathrm{c}=1$ and $L_\mathrm{c}=1$.
We derive expressions for excess length, entropy, and internal energy via (\ref{eq:14})-(\ref{eq:16}) 
from the function $B_{00}$ defined in (\ref{eq:12}), which in the current context, becomes
\begin{equation}\label{eq:f5} 
B_{00}= \sum_{m=0}^M\exp\Big(\beta\big[Jm^\nu-\gamma_m\big]\Big).
\end{equation}
For this quantity we have determined the leading term of an asymptotic expansion valid for tensions $1\ll J<J_M$, where $J_M\doteq M\nu/(1+\nu)+\nu-\frac{1}{2}$, including a first correction:
\begin{subequations}\label{eq:f6}
\begin{equation}\label{eq:f6a}
B_{00}\leadsto 
\exp\left(\frac{\beta}{\nu+1}\big(J-\nu+{\textstyle \frac{1}{2}}\big)\right)
b_\nu(\beta,J),
\end{equation}
\begin{equation}\label{eq:f6b}
b_\nu(\beta,J)=\sqrt{\frac{2\pi}{\beta\nu}}
\big(J-\nu+{\textstyle \frac{1}{2}}\big)^{(1-\nu)/2},
\end{equation}
\end{subequations}
where the factor (\ref{eq:f6b}) has been inferred from the integral
\begin{subequations}\label{eq:f7}
\begin{equation}\label{eq:f7a}
b_\nu(\beta,J)=\int_0^Mdt\,e^{-\beta f_\nu(J,t)},
\end{equation}
\begin{equation}\label{eq:f7b}
f_\nu(J,t)=\frac{1}{\nu+1}
\big(J-\nu+{\textstyle \frac{1}{2}}\big)^{\nu+1}+\gamma_t-Jt^\nu.
\end{equation}
\end{subequations}
The asymptotic results for the three quantities of interest thus become
\begin{equation}\label{eq:f8} 
\bar{L}\leadsto \bar{L}_\mathrm{as}=\big(J-\nu+{\textstyle \frac{1}{2}}\big)^{\nu}
+\frac{1-\nu}{2\beta(J-\nu+ \frac{1}{2})},
\end{equation}
\begin{equation}\label{eq:f9} 
\frac{\bar{S}}{k_\mathrm{B}}\leadsto\frac{\bar{S}_\mathrm{as}}{k_\mathrm{B}}=
\frac{1-\nu}{2}\ln\big(J-\nu+{\textstyle \frac{1}{2}}\big)
+\ln\sqrt{\frac{2\pi e}{\nu\beta}},
\end{equation}
\begin{align}\label{eq:f10} 
\bar{U}\leadsto\bar{U}_\mathrm{as}=&\frac{J\nu+\nu-\frac{1}{2}}{1+\nu}\big(J-\nu+{\textstyle \frac{1}{2}}\big)^\nu \nonumber \\
&\hspace{20mm}+\frac{J(2-\nu)-\nu+\frac{1}{2}}{2\beta(J-\nu+\frac{1}{2})}.
\end{align}

\begin{figure}[htb]
  \begin{center}
\includegraphics[width=40mm]{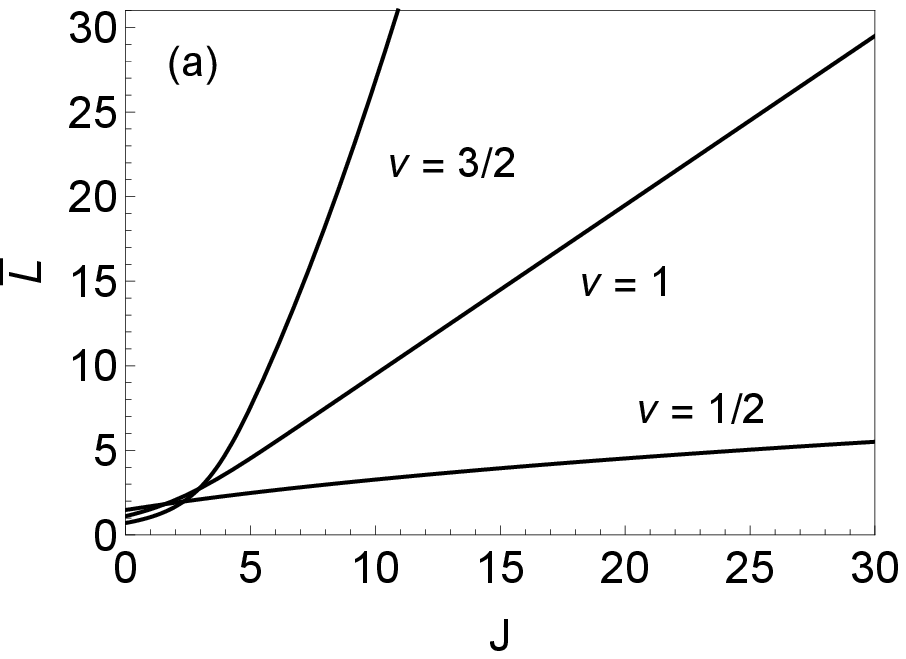}\hspace*{3mm}\includegraphics[width=40mm]{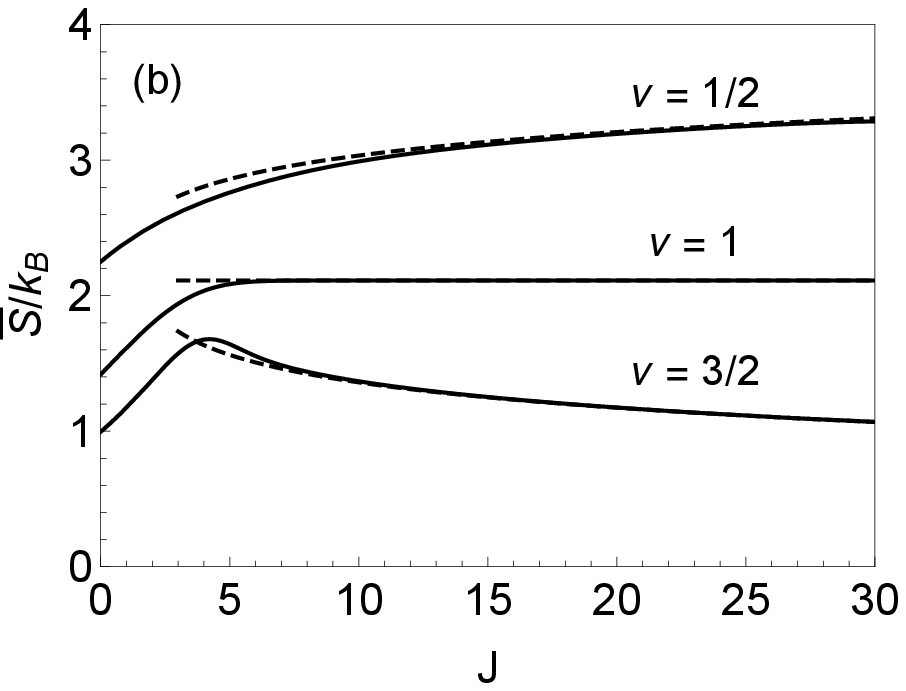}
\includegraphics[width=40mm]{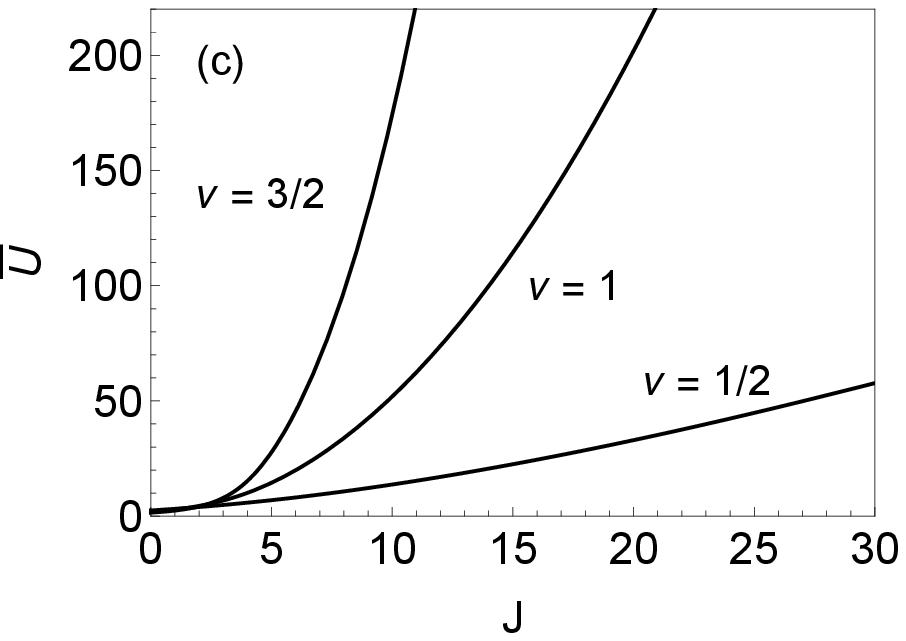}
\end{center}
\caption{(a) Scaled excess length, (b) entropy, and (c) internal energy, all versus scaled tension at constant scaled (inverse) temperature $\beta=0.25$, exponent 
$\nu=\frac{1}{2}, 1, \frac{3}{2}$, and $M=50$. 
The solid lines represent the results inferred from (\ref{eq:f5}). The asymptotic results (\ref{eq:f8})  and (\ref{eq:f10}) agree to within line thickness across the interval $5<J<30$. The asymptotic entropy result (\ref{eq:f9}) is shown dashed.}
  \label{fig:figure23}
\end{figure}

Exact results inferred from (\ref{eq:f5}) as explained in Sec.~\ref{sec:com-ext-par-1} are shown In Fig.~\ref{fig:figure23} for three exponent values, $\nu=\frac{1}{2}, 1, \frac{3}{2}$, and compared with the asymptotic results (\ref{eq:f8})-(\ref{eq:f10}). Across the range $5<J<30$ of scaled tension, the deviations are within the thickness of the lines except for the case of entropy.

\begin{figure}[b]
  \begin{center}
\includegraphics[width=40mm]{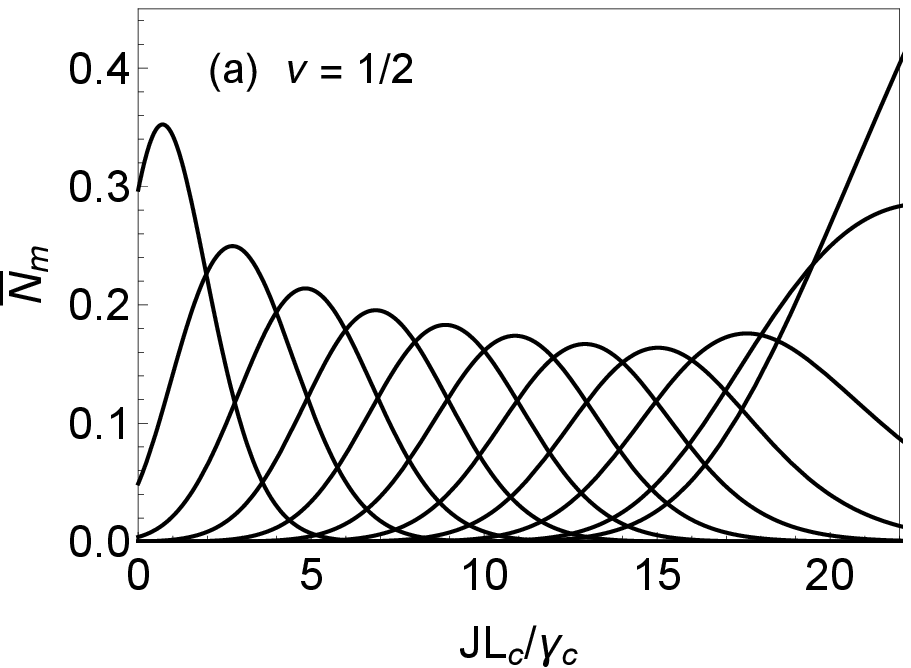}\hspace*{3mm}\includegraphics[width=40mm]{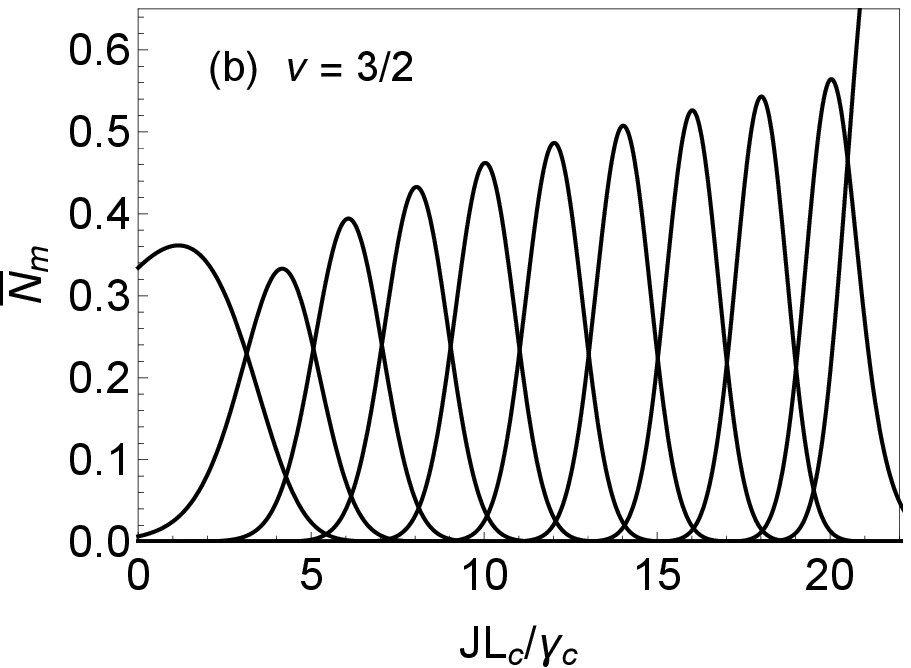}
\end{center}
\caption{Particle population densities $\bar{N}_m$, $m=1$, 3, 5, $\ldots$, 19, 20, (from left to right) versus scaled tension for two cases of power-law elasticity: (a) $\nu=\frac{1}{2}$ at $\beta=1.25$ and (b) $\nu=\frac{3}{2}$ at $\beta=0.3$. The case $\nu=1$ at $\beta=1$ is shown in Fig.~\ref{fig:figure5}(b).}
  \label{fig:figure8}
\end{figure}

The leading asymptotic term in the expressions for excess length and internal energy are power-laws with exponents $\nu$ and $\nu+1$, respectively. 
This term is independent of temperature in both expressions.
In the entropy expression, by contrast, temperature, plays a more prominent role.
The second term in the asymptotic result (\ref{eq:f9}), describes the plateau.
It depends on temperature but not on tension.
For $\nu=1$ this is the only term. 
The first term describes a deviation from the plateau value.
It is independent of temperature and increases (decreases) logarithmically with $J$ for $\nu<1$ ($\nu>1$).

We conclude this section with a `look under the hood'.
The population densities which produce linear elasticity come in an array of peaks of equal amplitude [Fig.~\ref{fig:figure5}(b)].
The power-law nonlinearity is produced by amplitude modulations of the kinds shown in Fig.~\ref{fig:figure8}.

%
\section{Entropic elasticity}\label{sec:appe}
%
At low tension, molecular chains are being straightened out in a process of entropic elasticity, named thermal unbending.
The \emph{freely jointed chain} (FJC) and \emph{worm-like chain} (WLC) models are widely used to describe that process quasistatically \cite{Flory69, Bust94, MS95, BWA+99, WQK15, FGWK12}.
We write $\bar{L}\doteq\langle L\rangle/L_0$ for the scaled length, where $L_0$ is the contour length.
For the scaled tension it is natural to use the dimensionless variable $\beta Jl_\mathrm{K}$ in the FJC model and $\beta Jl_\mathrm{p}$ in the WLC model, where the Kuhn segment length $l_\mathrm{K}$ and the persistence length $l_\mathrm{p}$ are different measures for bending stiffness.

\begin{figure}[htb]
  \begin{center}
\includegraphics[width=80mm]{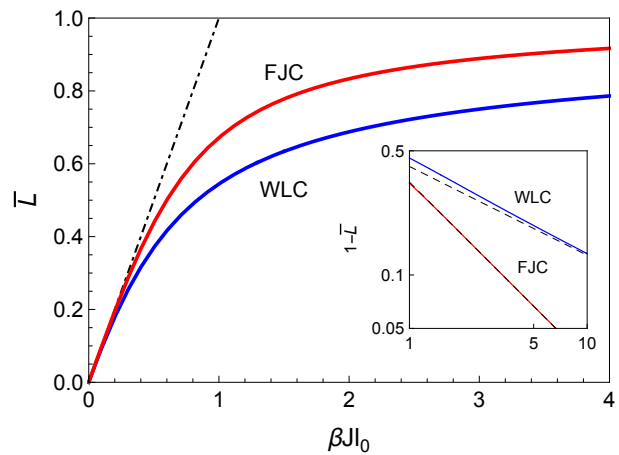}
\end{center}
\caption{Force-extension characteristics of the FJC and WLC models. The scaling lengths $l_0=\frac{1}{3}l_\mathrm{K}$ (FJC) and $l_0=\frac{2}{3}l_\mathrm{p}$ (WLC) chosen here produce a unit initial slope (dot-dashed line) for both models. The inset shows the deviation from saturation at strong tension in a log-log plot. The dashed lines have slopes $-1$ and $-\frac{1}{2}$.}
  \label{fig:figure7}
\end{figure}

The force-extension characteristics of the two models can thus be represented very concisely as
\begin{equation}\label{eq:e1} 
\bar{L}=\mathrm{coth}\big(\beta Jl_\mathrm{K}\big)-\frac{1}{\beta Jl_\mathrm{K}}\quad (\mathrm{FJC}),
\end{equation}
\begin{equation}\label{eq:e2} 
\beta Jl_\mathrm{p}=\bar{L}+\frac{1}{4}\left[\frac{1}{(1-\bar{L})^2}-1\right]\quad (\mathrm{WLC}),
\end{equation}
the last expression being a widely used and fairly accurate interpolation formula \cite{MS95}.
In Fig.~\ref{fig:figure7} we show graphical representations in two different formats.
The FJC chain is more compliant to unbending than the WLC chain. 
At strong tension the deviation from saturation approaches zero faster in the FJC model than in the WLC model, namely $\sim(\beta Jl_\mathrm{K})^{-1}$ as compared to $\sim(\beta Jl_\mathrm{p})^{-1/2}$.

\subsection{FJC thermal unbending}\label{sec:ther-unbe}
Here we show how the FJC result (\ref{eq:e1}) can be reproduced by an application of the compact level-1 compacts from Sec.~\ref{sec:com-ext-par-1}.
In this application they are contraction particles, contributing a negative extension upon activation.
Consider $M=2s$ species with combinatorial specifications (\ref{eq:10}).
A particle of species $m$, when thermally activated, contracts the molecular chain under tension $J$ through misalignment of a segment with Kuhn length $l_\mathrm{K}$ relative to the direction of tension.
For the amount of contraction we write,
\begin{equation}\label{eq:41} 
 \frac{\Delta L_m}{l_\mathrm{K}}=\frac{m}{s}\doteq 1-\cos\theta_m~:\quad m=1,\ldots,2s,
\end{equation}
where $\theta_m$ is the (discretized) polar angle of deviation.
The particle energy is equal to the contraction work associated with the activation,
\begin{equation}\label{eq:42} 
\epsilon_m=J\Delta L_m=Jl_\mathrm{K}\,\frac{m}{s}.
\end{equation}
The evaluation of the Gibbs free energy (\ref{eq:13}) with these activation energies yields the result,
\begin{equation}\label{eq:43} 
\beta \bar{G}=-\ln\left(\frac{\sinh\big(\beta Jl_\mathrm{K}(1+1/2s)\big)}
{\sinh\big(\beta Jl_\mathrm{K}/2s\big)}\right)+\beta Jl_\mathrm{K}.
\end{equation}
It is mathematically equivalent the free energy of a spin-$s$ Brillouin paramagnet.
The partial derivative (\ref{eq:14}) produces the force-extension characteristics,
\begin{align}\label{eq:44}
\bar{L}= & (1+1/2s)\coth\big(\beta Jl_\mathrm{K}(1+1/2s)\big) \nonumber \\
 &\hspace{15mm}-(1/2s)\coth\big(\beta Jl_\mathrm{K}/2s\big)-l_\mathrm{K}.
\end{align}
The state at infinite tension with all particles frozen out is the reference state in this application.
It has scaled contour length $l_\mathrm{K}$.
Taking the limit $s\to\infty$ and switching to the zero-tension reference state by adding $l_\mathrm{K}$ reproduces the FJC force-extension characteristic (\ref{eq:e1}).

\subsection{WLC thermal unbending}\label{sec:ther-unbe-wlc}
Reproducing the WLC force-extension characteristic via statistically interacting particles is more challenging but no less illuminating.
The comparison of the FJC and WLC force-extension characteristics in Fig.~\ref{fig:figure7} uses the relation,
\begin{equation}\label{eq:78} 
l_\mathrm{K}=2l_\mathrm{p},
\end{equation}
between Kuhn length and persistence length.
On this common length scale both models exhibit the same initial rise of extension under tension.
The internal bending stiffness, which is built into the WLC model only, manifests itself in the qualitatively slower approach to the high-tension asymptotic value of $\bar{L}$ (scaled contour length.)

We now combine two modules of our methodology previously introduced to produce a highly accurate representation of the WLC behavior.
We consider a set of $M_\mathrm{h}+1$ level-1 host species in combination with a set of $N_\mathrm{c}$ level-1 cap species.
The combinatorics and statistical mechanics of this kind of nested structure of particles is worked out in Appendix~\ref{sec:appg}.

The hosts $m=1,\ldots,M_\mathrm{h}$ represent extension particles with activation energies tailored in the manner explored in Sec.~\ref{sec:com-ext-par-1}, here to represent nonlinear elasticity which is entropic in part.
The caps $n=1,\ldots,N_\mathrm{c}$ represent contraction particles of the kind introduced in Sec.~\ref{sec:ther-unbe} for much the same purpose.
One cap can be activated from any host already activated.
One additional host, $m=0$, with a $J$-independent activation energy is needed to balance the effects of extension and contraction particles at zero tension.

Aiming for the specific purpose of reproducing the WLC force-extension characteristic at constant temperature, we have chosen the specifications,
\begin{subequations}\label{eq:79}
\begin{equation}\label{eq:79a} 
\beta\gamma_m^{(\mathrm{h})}=\frac{l_\mathrm{K}}{l_\mathrm{p}}
\sum_{m'=1}^{m-1}\left[\frac{1}{\sqrt{m'}}-\frac{1}{\sqrt{m}}\right],
\end{equation}
\begin{equation}\label{eq:79b}
L_m^{(\mathrm{h})}=2l_\mathrm{K}\left[1-\frac{1}{\sqrt{4m}}\right],
\end{equation}
\end{subequations}
for hosts $m=1,\ldots,M_\mathrm{h}$, and 
\begin{equation}\label{eq:80}
\beta\gamma_n^{(\mathrm{c})}=0,\quad 
L_n^{(\mathrm{c})}=-\frac{2l_\mathrm{K}}{N_\mathrm{c}}\,n,
\end{equation}
for caps $n=1,\ldots,N_\mathrm{c}$.
At room temperature, $\beta=0.244\mathrm{pN}^{-1}\mathrm{nm}^{-1}$, the host $m=0$ serves its intended purpose if we set $\beta\gamma_0^{(\mathrm{h})}=0.349$ and $L_0^{(\mathrm{h})}=0$ \cite{note2}.
The results presented below are for $l_\mathrm{K}=100$nm, $l_\mathrm{p}=50$nm.
With these specifications the fully extended chain has (scaled) length $\bar{L}=2l_\mathrm{K}$.

The free energy of this level-1 host-cap system is available in closed form for arbitrary numbers $M_\mathrm{h}$ and $N_\mathrm{c}$ [Appendix~\ref{sec:appg}]:
\begin{subequations}\label{eq:81}
\begin{align}\label{eq:81a}
& \beta\bar{G}=-\ln\left(1+Z_\mathrm{c}\sum_{m=0}^{M_\mathrm{h}}
e^{-\beta(\gamma_m^{(\mathrm{h})}-JL_m^{(\mathrm{h})})}\right), 
\\ \label{eq:81b}
& Z_\mathrm{c}=1+\sum_{n=1}^{N_\mathrm{c}}e^{\beta JL_n^{(\mathrm{c})}}.
\end{align}
\end{subequations}
Explicit results for the force-extension characteristic are derived from (\ref{eq:81}) as in previous applications.

\begin{figure}[htb]
  \begin{center}
\includegraphics[width=40mm]{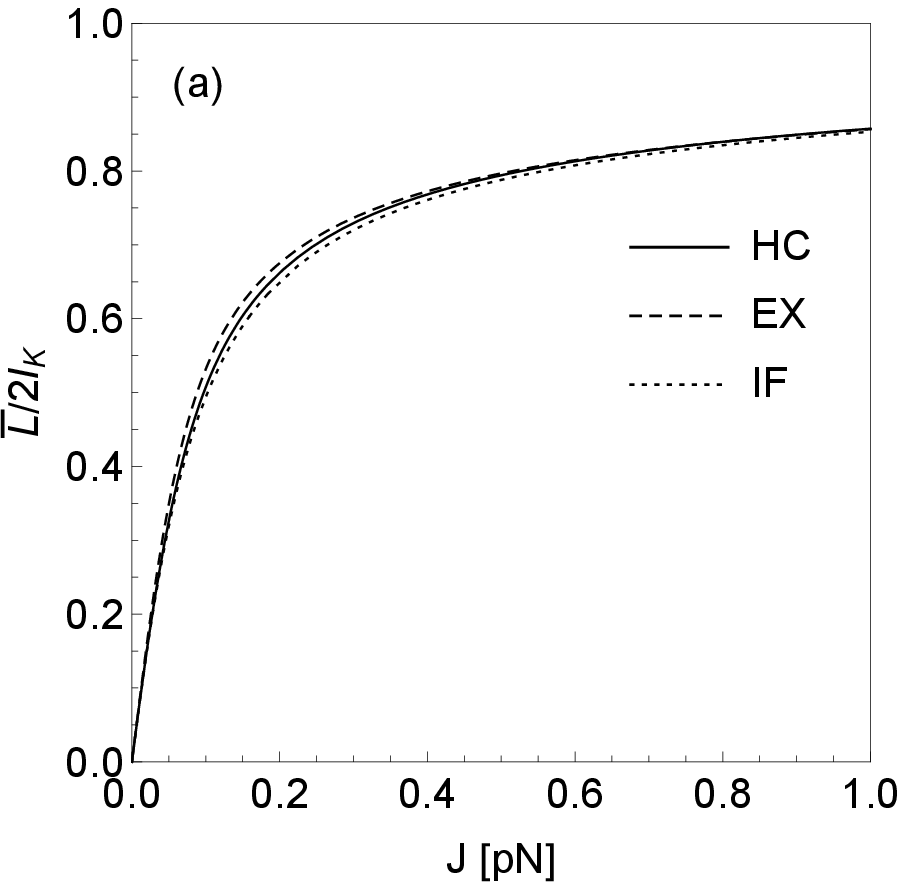}\hspace*{3mm}\includegraphics[width=40mm]{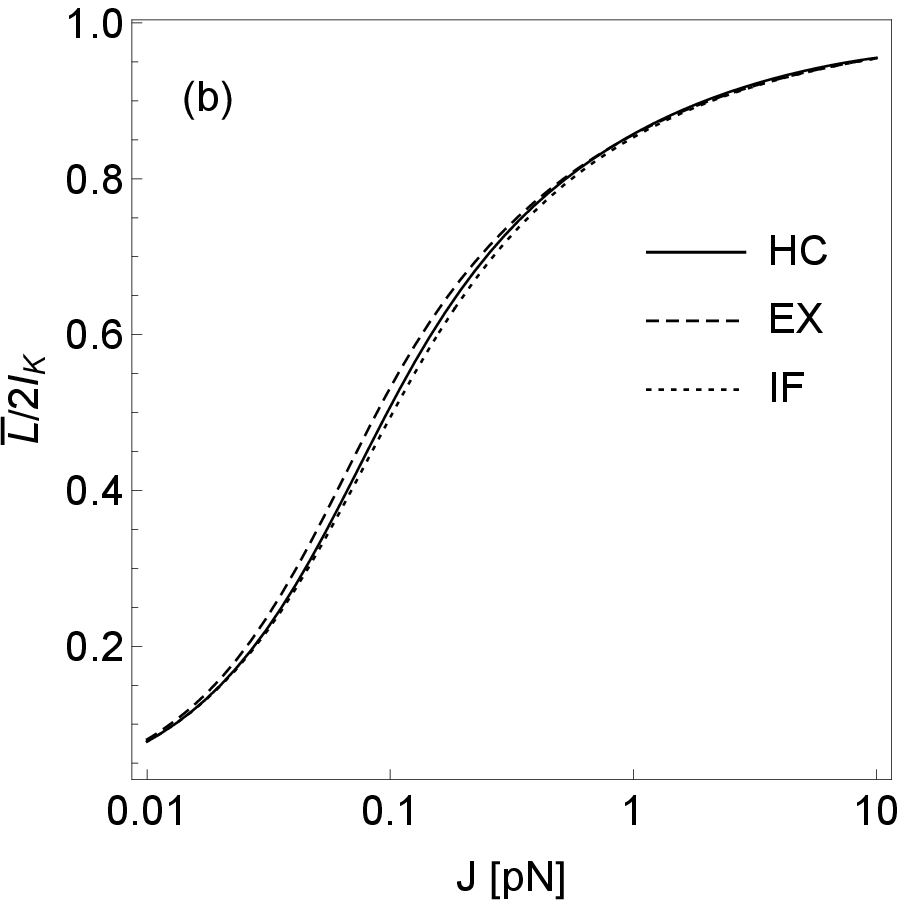}
\includegraphics[width=40mm]{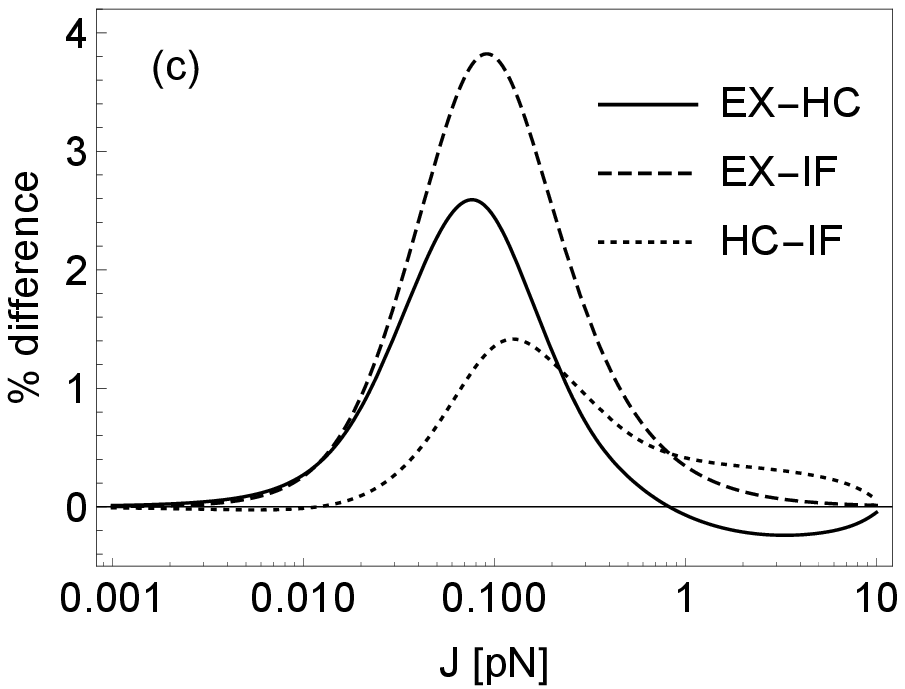}
\end{center}
\caption{Force extension characteristics in (a) linear-linear and (b) log-linear formats of the exact WLC model, named \textsf{EX} and taken from \cite{BWA+99}, the interpolation result (\ref{eq:e2}), named \textsf{IF}, and the host-cap result inferred from (\ref{eq:81}), named \textsf{HC} with $M_\mathrm{h}=200$ $N_\mathrm{c}=100$. (c) Percent difference between any two curves.}
  \label{fig:figure27}
\end{figure}

Graphical representations of $\bar{L}/2l_\mathrm{K}$ versus $J$ in two distinct formats are shown in Fig.~\ref{fig:figure27}.
For comparison we also show the interpolation result (\ref{eq:e2}) and the exact result taken from \cite{BWA+99}. 
Our result deviates from the exact result somewhat less than the interpolation result does.
The maximum deviation is below 3\%

There are different ways to model FJC and WLC behavior in the framework of this methodology. One alternative model, which employs level-2 nested particles (hosts, hybrids, and tags) is described in Appendix~\ref{sec:apph}.

%
\section{DNA under tension}\label{sec:DNA-1}
%
The force-extension characteristic of ds-DNA, mounted without torsional constraint, is well-known to exhibit distinct elastic responses in three successive regimes of applied tension \cite{MN13, Cluzel96, LRS+99, SAC+00, BSL+00, VMRW05, ZCF+12, ZCL+13}:
\begin{itemize}
\item[(i)] thermal unbending:~ $J ~\lesssim~ 10\mathrm{pN}$,
\item[(ii)] Hookean elasticity:~$10\mathrm{pN}\lesssim J ~\lesssim~ 65\mathrm{pN}$,
\item[(iii)] conformational change:~ $65\mathrm{pN}\lesssim J ~\lesssim~ 70\mathrm{pN}$.
\end{itemize}
The conformational change due to overstretching may represent a transition from B-DNA to S-DNA, DNA melting, or a combination of both \cite{Cluzel96, Smith96, BEP+12, BEP+14, KGB+13, BML+14}.
Here we demonstrate how regimes (i)-(iii) can be brought under one hat in the framework of the methodology developed thus far.

The foundation is laid by the  hosts and caps of Sec.~\ref{sec:ther-unbe-wlc} with energetic parametrizations (\ref{eq:79})-(\ref{eq:80}).
These particles assume the same functions here and describe the force-extension characteristic across regime (i).
For regime (ii) we use the particles introduced in Appendix~\ref{sec:appa} transposed from compacts to additional caps. 
It is straightforward to implement the switch of category [Appendix~\ref{sec:appg}].
The parametrization (\ref{eq:57}) of their energetics remains unchanged.
Regime (iii) requires one additional cap with activation energy in the form (\ref{eq:9}).

In this application, we are thus dealing with $M_\mathrm{h}+1$ hosts, each capable of hosting exactly one cap from $N_\mathrm{c}=N_\mathrm{c}^{(\mathrm{i})}+N_\mathrm{c}^{(\mathrm{ii})}+1$ species pertaining to the three regimes.
The free energy has the form (\ref{eq:81a}) with (\ref{eq:81b}) now modified to represent  three sets of cap species:
\begin{equation}\label{eq:58}
Z_\mathrm{c}=1+\sum_{n=1}^{N_\mathrm{c}^{(\mathrm{i})}}e^{\beta\epsilon_n^{(\mathrm{i})}}
+\sum_{l=1}^{N_\mathrm{c}^{(\mathrm{ii})}}e^{\beta\epsilon_l^{(\mathrm{ii})}}
+e^{\beta\epsilon^{(\mathrm{iii})}}.
\end{equation}
The specifications include characteristic lengths, tensions, and energies, six values in total:
\begin{align}\label{eq:59}
& \Delta L^{(\mathrm{i})}=110\mathrm{nm},~  \Delta L^{(\mathrm{ii})}=45.7\mathrm{nm}, ~\Delta L^{(\mathrm{iii})}=179\mathrm{nm}, \nonumber \\
& J_\mathrm{c}=79\mathrm{pN}, ~ \Delta J=75.9\mathrm{pN}, ~
 \gamma^{(\mathrm{iii})}=12105\mathrm{pNnm}.
\end{align}
Four of them are physical parameters, which have been known prior to their use here.
Only two are fitting parameters. 
$\Delta L^{(\mathrm{i})}$ is the widely used Kuhn's length of ds-DNA \cite{MN13, Bust94, cocco, Mark97}.  
$J_\mathrm{c}$ and $\Delta L^{(\mathrm{iii})}$ are well-established data for the overstretching transition \cite{Cluzel96, MN13, allemand, cocco, LRS+99}.
The two fitting parameters, $\Delta L^{(\mathrm{ii})}$ and $\Delta J$, pertain to the regime of contour elasticity.
The sixth parameter, $\gamma^{(\mathrm{iii})}$, is, in our modeling, a function of the other five parameters, too unwieldy to be stated here explicitly.
The integer values chosen in this application,
\begin{align}\label{eq:82}
M_\mathrm{h}=200,~ N_\mathrm{c}^{(\mathrm{i})}=60, ~
N_\mathrm{c}^{(\mathrm{ii})}=20.
\end{align}
merely control the smoothness of the result.
The activation energies which go into the free-energy expression (\ref{eq:81a}) with $Z_\mathrm{c}$ from (\ref{eq:58}) now have the following specifications, in part adaptations of (\ref{eq:79}), (\ref{eq:80}), and (\ref{eq:57}):
\begin{subequations}\label{eq:83}
\begin{align}\label{eq:83a}
& {\displaystyle \beta\gamma_m^{(\mathrm{h})}=2
\sum_{m'=1}^{m-1}\left[\frac{1}{\sqrt{m'}}-\frac{1}{\sqrt{m}}\right]}, \nonumber \\ 
&{\displaystyle L_m^{(\mathrm{h})}=2\Delta L^{(\mathrm{i})}\left[1-\frac{1}{\sqrt{4m}}\right]}
 \quad :~ m=1,\ldots M_\mathrm{h}, 
\end{align}
\begin{equation}\label{eq:83b}
\beta\gamma_n^{(\mathrm{i})}=0,\quad 
L_n^{(\mathrm{i})}=-\frac{2\Delta L^{(\mathrm{i})}}{N_\mathrm{c}^{(\mathrm{i})}}\,n\quad :~ 
n=1,\ldots,N_\mathrm{c}^{(\mathrm{i})},
\end{equation}
\begin{align}\label{eq:83c}
& \gamma_l^{(\mathrm{ii})}=\frac{l\Delta L^{(\mathrm{ii})}}{2N_\mathrm{c}^{(\mathrm{ii})}}
\left[\frac{l-1}{N_\mathrm{c}^{(\mathrm{ii})}}\,\Delta J+2(J_\mathrm{c}-\Delta J)\right], 
\nonumber \\
& L_l^{(\mathrm{ii})}=\frac{l}{N_\mathrm{c}^{(\mathrm{ii})}}\Delta L^{(\mathrm{ii})}\quad
:~ l=1,\ldots,N_\mathrm{c}^{(\mathrm{ii})},
\end{align}
\begin{equation}\label{eq:83d}
\epsilon^{(\mathrm{iii})}= \gamma^{(\mathrm{iii})}-J\Delta L^{(\mathrm{iii})}.
\end{equation}
\end{subequations}
The host $m=0$ balances extension against contraction at $J=0$, as in  Sec.~\ref{sec:ther-unbe-wlc}. Its elastic-energy constant is calculated analogously, yielding the value, $\beta\gamma_0^{(\mathrm{h})}=0.334$ with  $\beta=0.244\mathrm{pN}^{-1}\mathrm{nm}^{-1}$.

\begin{figure}[htb]
	\begin{center}
		\includegraphics[width=39mm]{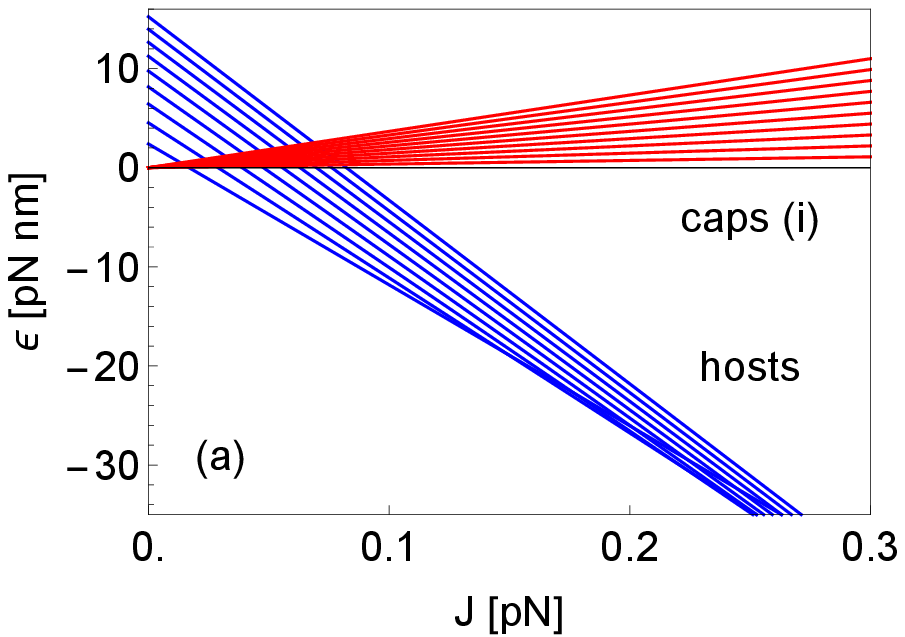}\hspace*{3mm}\includegraphics[width=42mm]{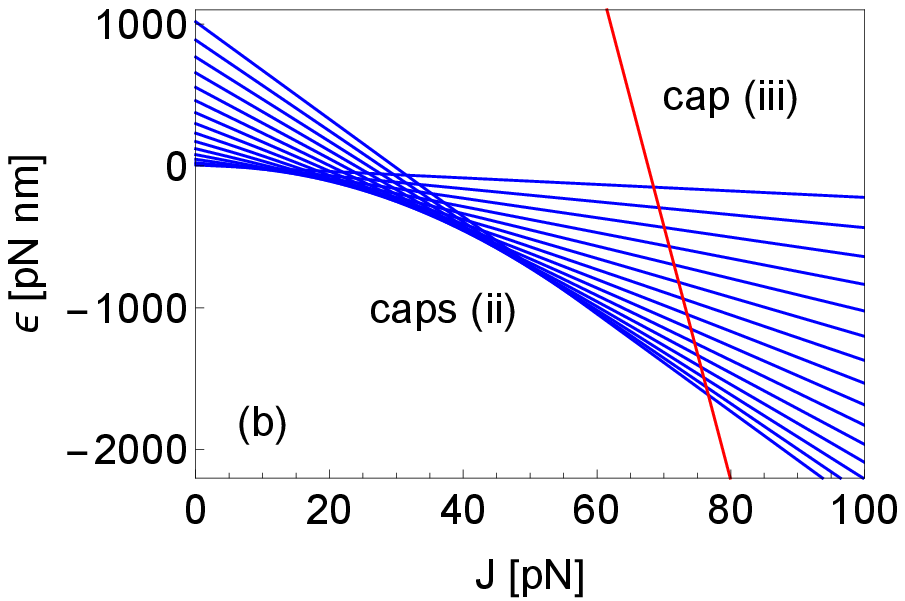}
	\end{center}
	\caption{Activation energies of (a) hosts for $m=2,\ldots,10$ and caps (i) for $n=1,\ldots,10$, (b) caps (ii) for $l=1,\ldots,15$ and cap (iii). Note the different scales left and right.}
	\label{fig:figure19}
\end{figure}

In Fig.~\ref{fig:figure19} we show the (linear) dependence on $J$ of the activation energies for some hosts and caps from sets (i), (ii), and (iii). 
Contraction particles have positive slope whereas extension particles have negative slope.
The lines representing hosts cross each other beyond the frame in a manner similar to caps (ii).

\begin{figure}[htb]
	\begin{center}
\includegraphics[width=82mm]{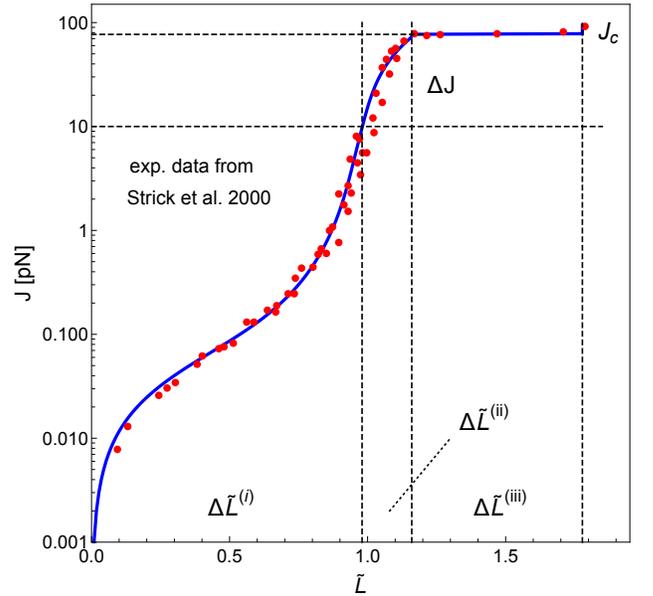}
	\end{center}
	\caption{Three-regime force-extension characteristic with specifications (\ref{eq:59})  with (absolute) tension $J$ on a logarithmic scale and length $\tilde{L}=L/L_0$ (scaled by contour length $L_0$) on a linear scale.
	The experimental data are from Ref.~\cite{SAC+00}.
	}
	\label{fig:figure20}
\end{figure}

The force-extension characteristic of torsionally unconstrained ds-DNA across all three regimes is derived from a single partition function.
Our result is displayed in Fig.~\ref{fig:figure20} on a logarithmic scale with the three regimes marked by vertical dashed lines.
It bears out a composite curve sketched two decades earlier \cite{Mark97}.
Overlaid in the same graph are experimental data compiled in Fig.~6A of Ref.~\cite{SAC+00}.
The agreement across four orders of magnitude of tension speaks for itself.

As $J$ increases from zero, the contraction particles, i.e. the caps (i), are gradually frozen out as their activation energies rise.
The system unbends gradually in consequence.
The WLC asymptotics of the unbending at $J\lesssim 10$pN is caused by the gradual population shift among hosts as their activation energies undergo crossings.
Once the contour length has been reached, entropic elasticity crosses over  into enthalpic elasticity.
Most bonds are populated by the host with the lowest energy. 
Now caps (ii) come into play gradually as their activation energies descend and undergo crossings with increasing tension, causing Hookean elongation beyond the contour length.
Linear elasticity terminates rather precipitously at $J\simeq65$pN, when the (steeply descending) activation energy of cap (iii) dips below that of the lowest cap (ii).
Here DNA undergoes a structural transformation, whose exact nature is still a matter of debate but for which there is overwhelming experimental evidence.

%
\section{Conclusion and outlook}\label{sec:concl}
%
In this work we have adapted a method of statistical mechanical analysis based on statistically interacting quasi-particles to the study of molecular chains under tension.
We have demonstrated its usefulness and versatility in a series of applications of increasing complexity.
These applications included instances of enthalpic elasticity (contour elongation), entropic elasticity (thermal unbending), and a structural transformation including effects of cooperativity.

One strength of this methodology is its modular nature.
Modules that describe specific types of elementary responses to tension can be combined in the design of a model for applications to molecular chains with complex force-extension characteristics.
This versatility has been demonstrated with an application to ds-DNA: the force-extension characteristic for torsionally unconstrained stretching across three regimes of qualitatively different elastic responses.

The power of this methodology for the study of elastic responses in molecular chains is far from exhausted by these applications. 
Two areas of extension currently in the works \cite{mct2,mct3} may be outlined as follows:
\begin{itemize}

\item Bonds between monomers or groups of monomers respond to any combination of tension and torque with the activation of a specific mix of statistically interacting particles carrying quanta of extension or contraction and quanta of twist or supercoiling.
Our methodology thus extended is capable of describing the conversion between twist chirality and plectonemic chirality.
It is also capable of reproducing the experimentally established phase diagram featuring  (native) B-DNA, (underwound) S-DNA, and (overwound) P-DNA \cite{MN13}.

\item Our methodology is amenable to extensions beyond quasistatic processes.
Such extensions will be explored mainly in the context of interactions between molecular chains and molecules of the embedding fluid and how these interaction affect the response to tension and torque.
Depending on their strength, the effects of intercalating bonds can be very diverse.
They may produce elastic softening or hardening due to electrostatic screening and hydrophobicity or they may result in effects of hysteresis or more general manifestations of irreversibility. 
One concrete goal is the modeling of the effect of the ethidium concentration on the DNA force-extension characteristic as experimentally observed (see Fig.~5(b) of Ref.~\cite{APRW16}.
Molecular docking for various purposes is another application of contact particles.

\end{itemize}

\acknowledgments
We thank DML Meyer for illuminating discussions on continuum mechanics.
We are grateful to J.-F. Allemand for the permission to use the data shown in Fig.~\ref{fig:figure20}.

\appendix

%
\section{Host with alternative caps}\label{sec:appg}
%
Here we amalgamate the analysis carried out in Sec.~\ref{sec:com-ext-par-1} for level-1 compacts and in Sec.~\ref{sec:nes-ext-par-1} for level-1 hosts and caps for applications in Secs~\ref{sec:ther-unbe-wlc} and \ref{sec:DNA-1}.
Consider a chain of $N$ monomers (with $N-1$ bonds).
Each bond can accommodate exactly one host (from $M_\mathrm{h}$ species).
Each host, when activated, can accommodate one or the other cap from $N_\mathrm{c}$ species.
The capacity constants for hosting and hosted particles are as explained earlier:
\begin{align}\label{eq:51} 
& A_m^{(\mathrm{h})}=N-1 \quad  :~ m=1,\ldots,M_\mathrm{h}, \nonumber \\
& A_n^{(\mathrm{c})}=0 \quad :~ n=1,\ldots,N_\mathrm{c}.
\end{align}

\begin{table}[b]
  \caption{Statistical interaction coefficients $g_{mm'}$ for $m<m'$ between two host species, $g_{mn}$ between a host and its hosted cap, $g_{nn'}$ for $n<n'$ between two cap species on the same host.}\label{tab:t2}
\begin{center}
\begin{tabular}{c|rc} \hline\hline 
 (i) & ~~$m$ & ~$m'$  \\ \hline 
$m$~ & $1$ & $1$ \\ 
$m'$~ & $0$ & $1$ \\ 
 \hline\hline
\end{tabular} \hspace{5mm}
\begin{tabular}{c|rr} \hline\hline 
(ii)  & ~~$m$ & ~~$n$  \\ \hline 
~$m$~ & $1$ & $0$ \\ 
$n$ & $-1$ & $1$ \\ 
 \hline\hline
\end{tabular} \hspace{5mm}
\begin{tabular}{c|rc} \hline\hline 
(iii) & ~~$n$ & ~$n'$  \\ \hline 
$n$~ & $1$ & $1$ \\ 
$n'$~ & $0$ & $1$ \\ 
 \hline\hline
\end{tabular}
\end{center}
\end{table}

The matrix of statistical interaction coefficients has $M_\mathrm{h}M_\mathrm{c}$ rows and columns.
All non-vanishing entries pertain to interactions of three kinds: 
(i) the statistical interaction between hosts is the same as that between compacts (Sec.~\ref{sec:com-ext-par-1});
 (ii) the statistical interaction between any host and its hosted cap is as previously stated (Sec.~\ref{sec:nes-ext-par-1});
(iii) the statistical interaction between alternative caps on the same host is again the same as that between compacts.
Caps on different hosts do not interact at all.
The interactions (i)-(iii) are restated summarily in Table~\ref{tab:t2}.

For the statistical mechanical analysis we set
\begin{align}\label{eq:52} 
& e^{\beta\epsilon_m^{(\mathrm{h})}}\doteq k_m^{(\mathrm{h})} 
\quad  :~ m=1,\ldots,M_\mathrm{h}, \nonumber \\
& e^{\beta\epsilon_n^{(\mathrm{c})}}\doteq k_n^{(\mathrm{c})} 
\quad :~ n=1,\ldots,N_\mathrm{c},
\end{align}
without specifying the activation energies $\epsilon_m^{(\mathrm{h})}$ and $\epsilon_n^{(\mathrm{c})}$ of hosts and caps, respectively.
The algebraic Eqs.~(\ref{eq:6}) can be solved separately for the caps. 
The solution, 
\begin{equation}\label{eq:53}
w_n^{(\mathrm{c})}=k_n^{(\mathrm{c})}\left[1+\sum_{n'=1}^{n-1}
\frac{1}{k_{n'}^{(\mathrm{c})}}\right],
\end{equation}
is akin to the solution (\ref{eq:11}) for compacts. 
The factor,
\begin{equation}\label{eq:54}
Z_\mathrm{c}\doteq \prod_{n=1}^{N_\mathrm{c}}
\frac{1+w_n^{(\mathrm{c})}}{w_n^{(\mathrm{c})}}=
1+\sum_{n=1}^{N_\mathrm{c}}\frac{1}{k_n^{(\mathrm{c})}},
\end{equation}
defined as the first expression and evaluated in the last expression using (\ref{eq:53}), appears in each of Eqs.~(\ref{eq:6}) for hosts. 
The solution,
\begin{equation}\label{eq:55}
w_m^{(\mathrm{h})}=\frac{k_m^{(\mathrm{h})}}{Z_\mathrm{c}}
\left[1+\sum_{m'=1}^{m-1}
\frac{Z_\mathrm{c}}{k_{m'}^{(\mathrm{h})}}\right],
\end{equation}
depends on the caps via that factor $Z_\mathrm{c}$, which then also appears in the partition function,
\begin{equation}\label{eq:56}
Z=\left[1+Z_\mathrm{c}\sum_{m=1}^{M_\mathrm{h}}
\frac{1}{k_m^{(\mathrm{h})}}\right]^{N-1}.
\end{equation}

%
\section{Regime of linear contour elasticity}\label{sec:appa}
\begin{figure}[htb]
	\begin{center}
\includegraphics[width=65mm]{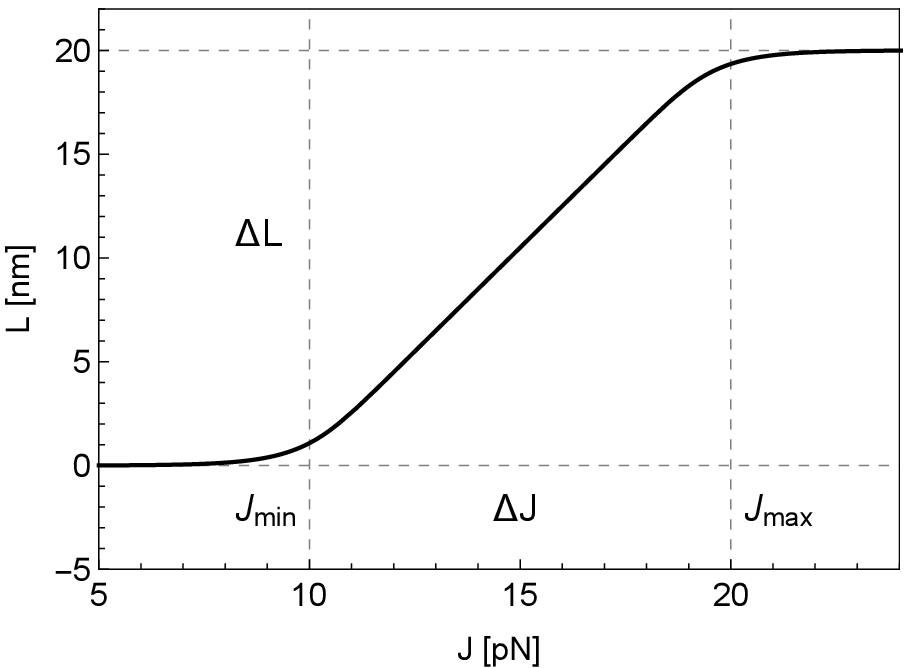}
	\end{center}
	\caption{Linear contour elangation vs tension for specifications $k_\mathrm{B}T=$1pNnm, $\Delta L=20$nm, $J_\mathrm{min}=\Delta J=10$nm, $M=20$.}
	\label{fig:figure9}
\end{figure}

%
It is useful to generalize the model of Hookean elasticity introduced in Sec.~\ref{sec:hooke} to a situation where it is applicable across a range $J_\mathrm{min}\leq J\leq J_\mathrm{min}+\Delta J$ of tension where it produces a contour extension of size $\Delta L$.
This generalization is straightforward and is being employed in Sec.~\ref{sec:DNA-1}.
The activation energies have the standard format (\ref{eq:9}), now with specifications,
\begin{equation}\label{eq:57}
\left. \begin{array}{l}
{\displaystyle L_m=\frac{m}{M}\Delta L} \\ \rule[-2mm]{0mm}{9mm} 
{\displaystyle \gamma_m=\frac{m\Delta L}{2M}
\left[\frac{m-1}{M}\Delta J+2J_\mathrm{min}\right]}
\end{array} \right\}~ :~ m=1,\ldots,M,
\end{equation}
that depend on the parameters $\Delta L$, $\Delta J$, $J_\mathrm{min}$ and the integer $M$.
The latter has no influence on the shape of the force-extension characteristic. Its value must be sufficiently high to guarantee smoothness at a given temperature.
One sample illustration is shown in Fig.~\ref{fig:figure9}.

%
\section{Nested level-2 particles}\label{sec:appb}
%
The level-2 hosts, hybrids, and tags previously introduced in the context of the coil-helix transition in polypeptides \cite{cohetra} are useful for applications to molecular chains under tension.
Here we present a somewhat generalized version of that model with an application to thermal unbending in mind [Appendix~\ref{sec:apph}].

From the reference state of $N$ monomers we nucleate segments of modified conformations by the activation of host particles as indicated in Fig.~\ref{fig:figure6}(f). 
These segments are then allowed to grow via the activation of hybrids and tags such that they contain a controllable amount of entropy as is expected in conformations that include some disorder.
The segments of modified conformation are best represented by a self-avoiding walk of controllable randomness in two dimensions as described in Ref.~\cite{cohetra}. 

The specifications for the multiplicity expression (\ref{eq:2}) are $M=2\mu$ with $\mu=1,2,\ldots$, $A_m=(N-2)\delta_{m,1}$, and nonzero interaction coefficients,
\begin{subequations}\label{eq:b1} 
\begin{equation}\label{eq:b1a} 
g_{1m'}=\left\{ \begin{array}{ll}
2 & :~ m'=1,\ldots,\mu, \\
1 & :~ m'=\mu+1,\ldots,2\mu,
\end{array} \right.
\end{equation}
\begin{equation}\label{eq:b1b} 
g_{mm'}=-1 ~:~\left\{ \begin{array}{l}
m'=m-1=1,\ldots,\mu-1, \\
m'=m-\mu+1=2,\ldots,\mu, \\
m'=m-\mu=1,\ldots,\mu.
\end{array} \right.
\end{equation}
\end{subequations}
The case $\mu=1$ has no hybrids.
It is used in Sec.~\ref{sec:coop} to describe cooperativity effects.
We are left with one host species $(m=1)$ and one tag species $(m=2)$.
The statistical interaction coefficients are $g_{11}=2$, $g_{12}=1$, $g_{21}=-1$, $g_{22}=0$.
The  coefficients for the case $\mu=3$ are tabulated in Table~\ref{tab:b1}.

\begin{table}[t]
  \caption{Statistical interaction coeffficients of the six species of quasiparticles that describe the case $\mu=3$.}\label{tab:b1} 
  \begin{tabular}{lc|rrrrrr}\hline\hline  \rule[-2mm]{0mm}{6mm}
 & $g_{mm'}$ & $1$ & $2$ & $3$ & $~~4$  & $~~5$ & $~~6$ \\ \hline \rule[-2mm]{0mm}{6mm}
host & $1$ & $2$ & $2$ & $2$ & $1$& $1$ & $1$\\ \rule[-2mm]{0mm}{5mm}
hybrid & $2$ & $-1$ & $0$ & $0$ & $0$ & $0$ & $0$\\  \rule[-2mm]{0mm}{5mm}
hybrid & $3$ & $0$ & $-1$ & $0$ & $0$ & $0$ & $0$\\ \rule[-2mm]{0mm}{5mm}
tag & $4$ & $-1$ & $-1$ & $0$ & $0$ & $0$ & $0$\\  \rule[-2mm]{0mm}{5mm}
tag & $5$ & $0$ & $-1$ & $-1$ & $0$ & $0$ & $0$\\ \rule[-2mm]{0mm}{5mm}
tag & $6$ & $0$ & $0$ & $-1$ & $0$ & $0$ & $0$\\ \hline\hline 
\end{tabular} 
\end{table} 

If we assign activation energies $\epsilon_\mathrm{n}$ to hosts, $\epsilon_\mathrm{h}$ to hybrids, and $\epsilon_\mathrm{g}$ to tags, we  have a four-parameter model: one discrete parameter $\mu$ and three continuous parameters $t,\tau,\vartheta$, 
\begin{equation}\label{eq:b2} 
\begin{array}{ll}
\mu=2,3,4,\ldots,\infty & \\
t\doteq e^{\beta\epsilon_\mathrm{h}/2} &:~ 0\leq t<\infty, \\
\tau\doteq e^{\beta(\epsilon_\mathrm{h}/2-\epsilon_\mathrm{n})} &:~ 0\leq\tau\leq 1, \\
\vartheta\doteq e^{\beta(\epsilon_\mathrm{h}/2-\epsilon_\mathrm{g})} &:~ 0\leq\vartheta\leq \infty. 
\end{array}
\end{equation}
In the context of the coil-helix transition worked out in Ref.~\cite{cohetra}, we have named $\mu$ range parameter, $t$ growth parameter, and $\tau$ nucleation parameter.
The value of what is now the fourth parameter was kept fixed at $\vartheta=1$.

The Gibbs free energy, from which most other quantities of interest are derivable, can be written in the form,
\begin{equation}\label{eq:b3} 
\bar{G} =-k_\mathrm{B}T\ln\big(1+w_{1}^{-1}\big), 
\end{equation}
where $w_1$ belongs to the set $w_m$, $m=1,\ldots,2\mu$ that solve the nonlinear algebraic Eqs.~(\ref{eq:6}). 
The physically relevant solution is reducible to a single polynomial equation of order $\mu+1$ for
\begin{equation}\label{eq:b4} 
\hat{w}\doteq\vartheta w,\quad w\doteq w_{\mu+1}(t,\tau,\vartheta),
\end{equation}
and recursive relations for the remaining $w_m$.
That polynomial equation is
\begin{equation}\label{eq:b5} 
(\vartheta+\hat{w}-t)S_{\mu}(\hat{w})=t\tau S_{\mu-1}(\hat{w}),
\end{equation}
where the $S_\mu(\hat{w})$ are Chebyshev polynomials of the second kind.
The recursive relations that complete the solution (for $\tau>0$) are of the form
 \begin{align} \label{eq:b6}
  w_{1} & = \frac{\hat{w}}{\tau}\frac{w_{2}}{1+w_{2}}, \nonumber \\
  w_{m} & = \hat{w}^{2} \frac{w_{m+1}}{1+w_{m+1}}-1
  \quad :~ m=2,...,\mu-1, \nonumber  \\ 
w_{\mu} & = \hat{w}^{2}-1\quad :~ w_{\mu+1} =\cdots=w_{2\mu}=w.
\end{align}

A singularity emerges in the solution of (\ref{eq:b5}) for ${\mu\to\infty}$ at $\tau>0$.
The transition occurs at
\begin{equation}\label{eq:b8} 
t_\mathrm{c}\doteq \frac{2+\vartheta}{1+\tau}
\end{equation}
and the analytic solution has the form
\begin{equation}\label{eq:b9}
\hat{w}=\left\{ \begin{array}{ll} 2 & : 0\leq t\leq t_\mathrm{c}, \\ 
t-\vartheta+{\displaystyle \frac{t\tau}{\lambda}} & : t>t_\mathrm{c},
\end{array} \right. 
\end{equation}
\begin{equation}\label{eq:b10} 
\lambda\doteq \frac{1}{2}\left[t-\vartheta+\sqrt{(t-\vartheta)^2+4(1-t\tau)}\,\right].
\end{equation}
It follows that the Gibbs free energy (\ref{eq:b3}) can be evaluated with
\begin{equation}\label{eq:b11} 
w_1=\left\{ \begin{array}{ll}
{\displaystyle \frac{t}{2+\vartheta-t}} & :~ t<t_\mathrm{c}, \\ \rule[-2mm]{0mm}{8mm}
{\displaystyle \frac{\lambda}{\tau}} & :~ t\geq t_\mathrm{c},
\end{array} \right.
\end{equation}
inferred from (\ref{eq:b9}).

%
\section{FJC and WLC from nested particles}\label{sec:apph}
%
Here we explore the versatility of the level-2 hosts, hybrids, and tags introduced in Appendix~\ref{sec:appb} for the purpose of simulating the FJC and WLC force-extension relations.
We set $\mu=\infty$ and $\tau=1$.
This leaves the two parameters $t$ and $\vartheta$.
We begin by declaring the tags to be extension particles, implying that their activation energy (in units of $\beta^{-1}$) depends linearly on tension with negative  slope.
We thus write
\begin{equation}\label{eq:e3} 
\beta\epsilon_\mathrm{g}=-\beta aJl_\mathrm{ch}
\end{equation}
with $0<a<1$ representing the bond length in units of the characteristic length $l_\mathrm{ch}$, which will stand for the Kuhn length $l_\mathrm{K}$ in the FJC application and for the persistence length $l_\mathrm{p}$ in the WLC application.
The role of the hosts and hybrids is to be discovered from the $J$-dependence of their activation energy.

The model specifications are encoded in the two functions $t(\beta Jl_\mathrm{ch})$ and $\vartheta(\beta Jl_\mathrm{ch})$.
They are constrained by the activation energy (\ref{eq:e3}) in the form
\begin{equation}\label{eq:e4} 
\frac{t(x)}{\vartheta(x)}=e^{\beta\epsilon_\mathrm{g}(x)}=
e^{-ax},\quad x\doteq\beta Jl_\mathrm{ch}
\end{equation}
and determine the activation energy $\beta\epsilon_\mathrm{h}(x)$ via
\begin{equation}\label{eq:e5} 
t(x)=e^{\beta\epsilon_\mathrm{h}(x)/2}.
\end{equation}

Next we must determine on which side of the critical point (\ref{eq:b8}) in the $(t,\vartheta)$ parameter plane our FJC and WLC applications are located. 
As it turns out, the inequality
\begin{equation}\label{eq:e6}
t(x)<t_\mathrm{c}(x)=1+\frac{\vartheta(x)}{2},
\end{equation}
is always satisfied.
It then follows from expression (\ref{eq:b3}) for the Gibbs free energy and the applicable solution (\ref{eq:b9}) that the force-extension relation, inferred via derivative as in (\ref{eq:14}), depends on $t(x)$ and $\vartheta(x)$ as follows:
\begin{align}\label{eq:e7}
\bar{L}(x) &=\frac{\vartheta'(x)}{2+\vartheta(x)}
-\frac{t'(x)}{t(x)} \nonumber \\
&=\frac{\vartheta'(x)}{2+\vartheta(x)}
-\frac{\vartheta'(x)}{\vartheta(x)}+a,
\end{align}
where we have used (\ref{eq:e4}) to arrive at the last expression.

In applications to empirical force-extension relations $\bar{L}(x)$ such as (\ref{eq:e1}) or (\ref{eq:e2}) we can integrate (\ref{eq:e7}).
The general expression reads,
\begin{equation}\label{eq:e11} 
\vartheta(x)=2\left[\displaystyle 3e^{-ax}\int_0^xdx'\bar{L}(x')-1\right]^{-1}.
\end{equation}
The initial condition, $\vartheta(0)=1$, guarantees that all activation energies vanish at zero tension.
This choice is not mandatory but it simplifies the interpretation.
The solution (\ref{eq:e11}) then solely depends on the parameter $a$.

\begin{figure}[htb]
  \begin{center}
\includegraphics[width=40mm]{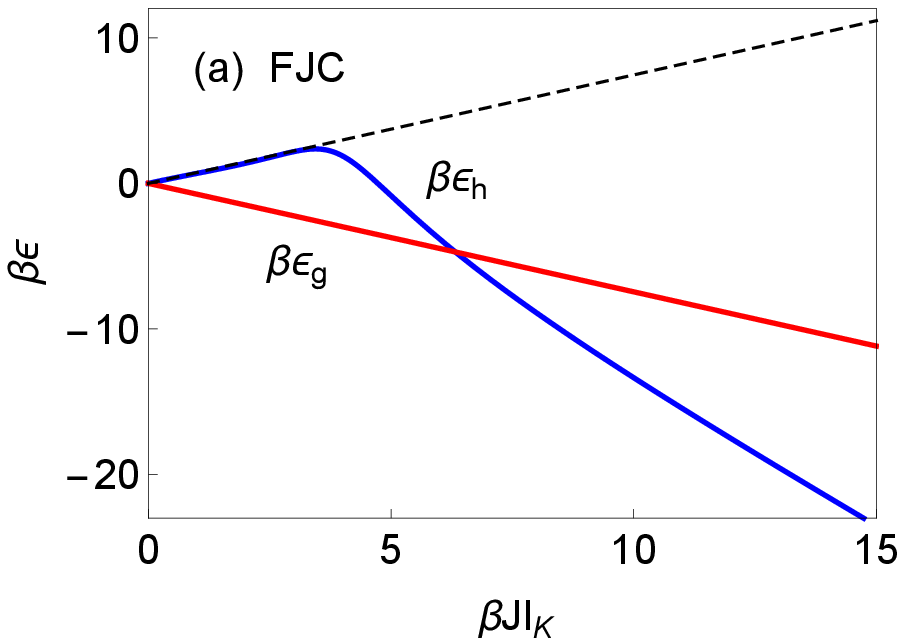}\hspace*{3mm}\includegraphics[width=40mm]{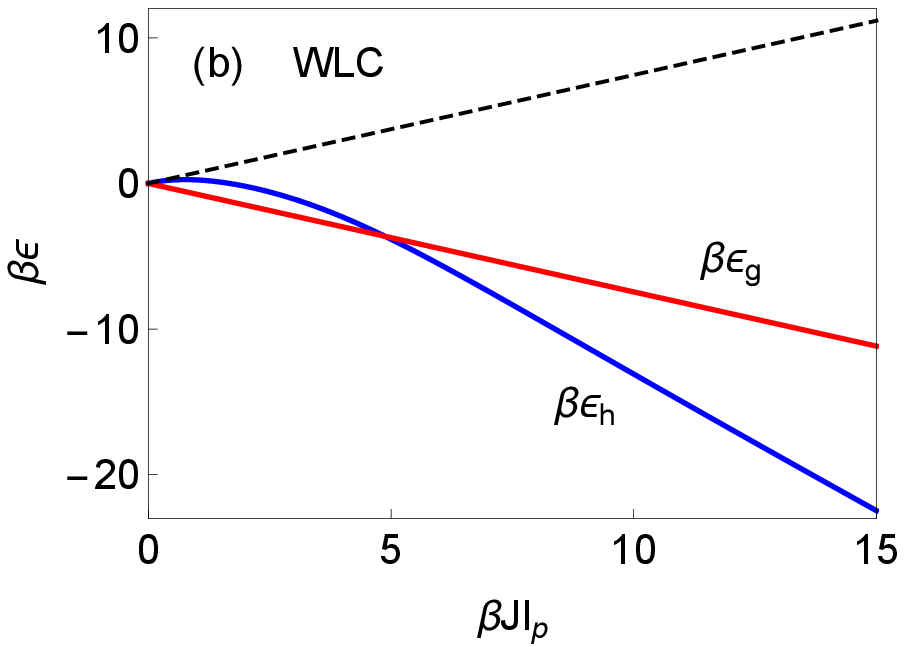}
\end{center}
\caption{Scaled activation energies versus scaled tension that reproduce the force-extension characteristics of (a) the FJC model and (b) the WLC model (interpolation formula). The parameter value, $a=0.745$ has been used for both models. The dashed and solid straight lines have slope $\pm a$. The curved line has asymptotic slope $-2$.}
  \label{fig:figure26}
\end{figure}

In the case of the  FJC model expression (\ref{eq:e11}) with $\bar{L}(x)$ from (\ref{eq:e1}) becomes,
\begin{equation}\label{eq:e8}
\vartheta(x)=t(x)e^{ax}=\frac{2x}{3\sinh(x)e^{-ax}-x}.
\end{equation}
Here $a<1$ is the bond length in units of the Kuhn length $l_\mathrm{K}$.
The particle activation energies extracted from this solution via (\ref{eq:e4}) and (\ref{eq:e5}) are plotted versus scaled tension in Fig.~\ref{fig:figure26}(a).

By design, the tags are extension particles and have activation energy $\beta\epsilon_\mathrm{g}$ which decrease from zero linearly with tension. 
The activation energy $\beta\epsilon_\mathrm{h}$ of hybrids is reverse-engineered to produce the force-extension characteristic (\ref{eq:e1}).
We see that its dependence on tension consists of two close to linear stretches, one with positive slope and the other with negative slope.

At low to moderate tension, $\beta Jl_\mathrm{K}\lesssim 4$, the hybrids act the part of contraction particles.
In the zero-tension limit, where all particles have vanishing activation energy, the thermally activated contraction particles exactly counteract the effect of the thermally activated extension particles.
With tension increasing from zero, the population of contraction particles is suppressed on account of their rising activation energy whereas the population density of extension particles is enhanced as their activation energy goes negative.
The almost linear variations of both activation energies produces the FJC force-extension curve through the first third of the interval shown in Fig.~\ref{fig:figure7}.

At higher tension, $\beta Jl_\mathrm{K}\gtrsim 4$, the hybrids assume a different role, namely that of extension particles with a nonzero elastic-energy constant.
With tension increasing from the crossover point, the population of the original extension particles are gradually crowded out by contraction particles turned into  extension particles. 
The full range of tension $0<\beta Jl_0<4$ shown in Fig.~\ref{fig:figure7} is equivalent to the range $0<\beta Jl_\mathrm{K}<12$ in Fig.~\ref{fig:figure26}(a).
The asymptotic regime begins at $\beta Jl_\mathrm{K}\gtrsim7$, where the asymptotic slope of the hybrid activation energies sets in. 

In the case of the WLC model we have to resort to a numerical analysis of the integral in (\ref{eq:e11}) with the physically relevant solution of the cubic equation (\ref{eq:e2}) used for $\bar{L}(x)$.
The results are shown in Fig.~\ref{fig:figure26}(b).
Here $a<1$ is the bond length in units of the persistence length $l_\mathrm{p}$.
We see similarities and differences between the two models.
Here our interpretation of the role of hybrids, which depends on a linear $J$-dependence of their activation energies, is limited to high tension.

The range of tension covered in Fig.~\ref{fig:figure26}(b) is more than twice as wide  than the range covered in Fig.~\ref{fig:figure7} for the WLC curve: $0<\beta Jl_0<4$ is equivalent to $0<\beta Jl_\mathrm{p}<6$.
It is evident that the hybrid activation energy straightens out to represent true extension particles at high tension.
The asymptotic slope of the hybrid particles is the same as in the FJC case.
The integral in (\ref{eq:e11}) with the asymptotic WLC extension characteristic,
\begin{equation}\label{eq:e9}
\bar{L}(x)_\mathrm{as}=1-\frac{1}{2\sqrt{x}},
\end{equation}
produces the analytic solution,
\begin{equation}\label{eq:e10}
\vartheta(x)_\mathrm{as}=\frac{2e^{2b+\sqrt{x}+(a-1)x}}
{1-e^{2b+\sqrt{x}+(a-1)x}}.
\end{equation}
The numerical solution inferred from the the full WLC force-extension relation (\ref{eq:e2}) with $a=0.745$ is matched near perfectly for $\beta Jl_\mathrm{p}\gtrsim 7$ by the asymptotic solution (\ref{eq:e10}) if we set $b=-0.202$. 

The remarkable fact is that both the FJC and WLC asymptotic regimes can be simulated, within the framework of the mathematical model used here, as the interplay of two kinds of extension particles with identical asymptotic slopes in their $J$-dependence.




\begin{thebibliography}{100}
\bibitem{note2}
The value quoted in the text is derived from the expression, $e^{-\beta\gamma_0^{(\mathrm{h})}}=\sum_{m=1}^{M_\mathrm{h}}[L_m^{(\mathrm{h})}/l_\mathrm{K}-1]e^{-\beta\gamma_m^{(\mathrm{h})}}$.
\end{thebibliography}
\end{document}